\documentclass[twocolumn]{aastex631}

\usepackage{graphicx}
\usepackage{amsmath}
\usepackage{amssymb}
\usepackage{newtxtext,newtxmath}
\usepackage{hyperref}
\usepackage{gensymb}
\usepackage{enumitem}
\usepackage{booktabs}

\newcommand{\Msun}{\ensuremath{M_{\odot}}}
\newcommand{\Gyr}{\ensuremath{\textrm{Gyr}}}
\newcommand{\Myr}{\ensuremath{\textrm{Myr}}}

\newcommand{\kpc}{\ensuremath{\textrm{kpc}}}

\newcommand{\pc}{\ensuremath{\textrm{pc}}}

\newcommand{\FeH}{\ensuremath{[\textrm{Fe}/\textrm{H}]}}
\newcommand{\MgFe}{\ensuremath{[\textrm{Mg}/\textrm{Fe}]}}

\newcommand{\alphaFe}{\ensuremath{[\alpha/\textrm{Fe}]}}

\newcommand{\dex}{\ensuremath{\textrm{dex}}}
\newcommand{\Msunyr}{\ensuremath{\Msun/\textrm{yr}}}
\newcommand{\Msunyrdex}{\ensuremath{\Msun/\textrm{yr}/\textrm{dex}}}

\newcommand{\rhalf}{\ensuremath{r_{\star,\textrm{half}}}}
\newcommand{\SFR}{\ensuremath{\textrm{SFR}}}

\begin{document}

\title{Rising from the Ashes II: The Bar-driven Abundance Bimodality of the Milky Way}

\author{Angus Beane}
\affiliation{Center for Astrophysics $|$ Harvard \& Smithsonian, Cambridge, MA, USA}

\author{James W. Johnson}
\affiliation{Carnegie Science Observatories, Pasadena, CA, USA}

\author{Vadim A. Semenov}
\affiliation{Center for Astrophysics $|$ Harvard \& Smithsonian, Cambridge, MA, USA}

\author{Lars Hernquist}
\affiliation{Center for Astrophysics $|$ Harvard \& Smithsonian, Cambridge, MA, USA}

\author{Vedant Chandra}
\affiliation{Center for Astrophysics $|$ Harvard \& Smithsonian, Cambridge, MA, USA}

\author{Charlie Conroy}
\affiliation{Center for Astrophysics $|$ Harvard \& Smithsonian, Cambridge, MA, USA}

\begin{abstract}
    The Milky Way hosts at least two modes in its present day distribution of Fe and $\alpha$-elements. The exact cause of this bimodality is disputed, but one class of explanations involves the merger between the Milky Way and a relatively massive satellite (Gaia-Sausage-Enceladus) at $z\sim2$. However, reproducing this bimodality in simulations is not straightforward, with conflicting results on the prevalence, morphology, and mechanism behind multimodality. We present a case study of a galaxy in the Illustris TNG50 simulation which undergoes sequential phases of starburst, brief quiescence, and then rejuvenation. This scenario results in a pronounced abundance bimodality after a post-processing adjustment of the \alphaFe{} of old stars designed to mimic a higher star formation efficiency in dense gas. The high- and low-$\alpha$ sequences are separated in time by the brief quiescent period, which is not associated with a merger but by the formation of a bar followed by AGN activity. This galaxy indicates a novel scenario in which the $\alpha$-bimodality in the Milky Way is caused by the formation of the bar via AGN-induced quenching. In addition to a stellar age gap in the Milky Way, we predict that abundance bimodalities should be more common in barred as opposed to unbarred galaxies.
  \end{abstract}
    
  \keywords{Classical Novae (251) --- Ultraviolet astronomy(1736) --- History of astronomy(1868) --- Interdisciplinary astronomy(804)}

\section{Introduction}\label{sec:intro}
The stellar surface abundances of most elements retain the composition of their natal gas cloud. Therefore, the present-day distribution of their abundances encodes the chemical history of a galaxy's gas phase. Two classes of elements have received particular attention in the Milky Way due to their disparate formation channels: iron-peak (such as Fe) and $\alpha$-elements (produced through the $\alpha$-process, such as O and Mg). Enrichment of iron-peak elements is primarily through Type~Ia and Type~II supernovae (SNe), whereas enrichment of $\alpha$-elements is primarily only through Type~II SNe.

The bimodality inherits a long history of attempts to decompose the disk dating back to \citet{1983MNRAS.202.1025G}, who showed that the vertical distribution of stars is well-fit by a double exponential. This led to a separation based on kinematics between the ``thin'' and ``thick'' disk. It was later shown that the thick disk is more $\alpha$-enhanced than the thin disk \citep{1996ASPC...92..307G,1998A&A...338..161F,2004AN....325....3F,2006MNRAS.367.1329R}. However, it was not until more recently that it became obvious there is a clean separation between two sequences in the abundance plane, i.e., the $\alpha$-rich and $\alpha$-poor disk \citep{2011A&A...535L..11A,2012A&A...545A..32A,2014A&A...562A..71B,2014ApJ...796...38N,2020MNRAS.493.2952H}, also shown in the upper left panel of Figure~\ref{fig:fig1}. The Milky Way is the only galaxy for which such a structure has been definitively shown to exist.\footnote{Note claims of a detection \citep{2023ApJ...956L..14K} and nondetection \citep{2024IAUS..377..115N} in M31.}

The distinction between a chemical and kinematic decomposition of the disk is illustrated by a population of low-$\alpha$ stars on vertically extended orbits in the outer disk \citep{2015ApJ...808..132H,2016ApJ...823...30B}. This is thought to arise from inside-out formation \citep[which predict and claim a negative age gradient at high altitude]{2015ApJ...804L...9M,2016ApJ...831..139M}, or from other processes, including misaligned gas accretion and minor mergers \citep{2010MNRAS.408..783R,2009MNRAS.396..696S}. Nonetheless, in the outer disk, the kinematically-defined thick disk has contributions from both the chemically defined high- and low-$\alpha$ sequences.

The origin of the bimodality is a topic of active debate, with three main scenarios proposed to explain it. First, it is a result of internal secular processes that generate the bimodality through radial migration \citep{2009MNRAS.396..203S,2021MNRAS.507.5882S,2023MNRAS.523.3791C} or clump formation \citep{2019MNRAS.484.3476C,2020MNRAS.492.4716B,2021MNRAS.502..260B,2023ApJ...953..128G}. Second, the bimodality is generated through gas infall scenarios, either from specific gas accretion episodes from the intergalactic medium \citep{1997ApJ...477..765C,2009IAUS..254..191C,2017MNRAS.472.3637G,2019A&A...623A..60S}, or through a more self-consistent collapse sequence of the circumgalactic medium driven through feedback \citep{2021MNRAS.501.5176K}. Third and finally, the bimodality is generated through a merger process, either by enhancing the star formation rate (SFR) of the Galaxy \citep{2018MNRAS.474.3629G}\footnote{\citet{2004ApJ...612..894B,2005ApJ...630..298B,2007ApJ...658...60B,2010MNRAS.402.1489R} also explored the $\alpha$-enhancement of the thick disk resulting from gas-rich mergers. However, they did not explore the arising of a clean separation purely in chemistry.} or by supplying a relatively pristine gas supply that resets the metallicity of the Galaxy \citep{2020MNRAS.491.5435B,2024MNRAS.528L.122C}.

Strong evidence that the Milky Way did undergo a merger with the Gaia-Sausage-Enceladus (GSE) satellite supports the merger-related scenarios \citep{2018MNRAS.478..611B,2018Natur.563...85H,2020ApJ...901...48N,2024ApJ...972..112C}. In deriving stellar birth radii from assuming a linear relation between \FeH{} and radius with a time-evolving slope, \citet{2024MNRAS.535..392L} argued that the bimodality resulted from a steepening of the metallicity gradient at the time of the GSE merger \citep[see also][]{2023MNRAS.525.2208R}. This provides further evidence that the bimodality formed around the same time as the GSE merger, although the case for a causal connection is less clear -- the bar seemed to form around the same time \citep[e.g][]{2019MNRAS.490.4740B,2024MNRAS.530.2972S}.

In \citet{2024arXiv240707985B} (hereafter Paper~I), we proposed an alternate scenario for the formation of the bimodality, driven by a brief ($\sim300\,\Myr$) quiescent period in the Galaxy's history in a narrow metallicity bin, assuming the gas-phase \alphaFe{} is declining sufficiently rapidly.\footnote{An observational claim with a longer period of global quiescence was made in \citet{2016A&A...589A..66H}.} The lower SFR results in fewer stars forming in the intermediate region between the high- and low-$\alpha$ sequences, reducing the occurrence of this transitional population in present-day observations. A global quiescent period is sufficient for producing such gaps in the metallicity-dependent SFR, though it is not necessary. This mechanism resembles two-phase infall models that incorporate a temporary halt in star formation \citep[][and references therein]{2024arXiv240511025S}, though the quiescent period in Paper~I is much shorter.

In Paper~I, we used idealized simulations of a galaxy merger that triggered the quiescent period. However, in that work we argued that it was the quiescent period, not the merger, which was necessary to produce a bimodality. To demonstrate this, here we study a galaxy from the Illustris TNG50 cosmological simulation. This galaxy exhibits the sequence of events presented in Paper~I after a post-processing step, motivated by recent work showing that the star formation efficiency (SFE) of dense gas at high-$z$ is too low in the TNG model \citep{2024arXiv240909121H}, which increases the \alphaFe{} of old star particles. The galaxy undergoes a brief quiescent period which neatly separates a high- and low-$\alpha$ sequence. However, instead of being preceded by a merger, the quenching is preceded by apparent bar-induced AGN activity. Therefore, this work serves as a verification that the scenario in Paper~I is possible in cosmological simulations and does not require a merger.

In Section~\ref{sec:methods}, we discuss our selection technique which led to discovering the galaxy, the observations we use for comparison, as well as a simple one zone chemical evolution model we use to justify our post-processing step. In Section~\ref{sec:results}, we present the main results which we discuss and interpret in Section~\ref{sec:disc}. We conclude in Section~\ref{sec:conc}.

\section{Methods}\label{sec:methods}
\subsection{IllustrisTNG Sample}\label{ssec:tng}
We use Illustris TNG50 \citep{2019MNRAS.490.3196P, 2019MNRAS.490.3234N, 2019ComAC...6....2N}, a cosmological simulation of a $\sim50\,\textrm{cMpc}$ box at high resolution ($m_{\textrm{baryon}}\sim8.5\times10^4\,\Msun$). It uses the gravito-magneto-hydrodynamics code \texttt{AREPO} \citep{2010MNRAS.401..791S, 2016MNRAS.455.1134P}, along with the TNG galaxy formation model \citep{2013MNRAS.436.3031V, 2017MNRAS.465.3291W, 2018MNRAS.473.4077P}. This model includes several subgrid processes, including a wind generation model, chemical enrichment from SNe and asymptotic giant branch stars, and thermal and kinetic feedback from AGN.

There are two pieces of the TNG model of note for this work. First is the black hole (BH) accretion and feedback method \citep{2017MNRAS.465.3291W}. The BH accretion rate ($\dot{M}_{\textrm{BH}}$) is computed using the local structure of the gas phase with the Bondi-Hoyle-Lyttleton formula \citep{1939PCPS...35..405H,1944MNRAS.104..273B,1952MNRAS.112..195B}, with a maximum of the Eddington accretion rate ($\dot{M}_{\textrm{edd}}$). The model allows the BH to be either in a kinetic radio-mode or a thermal quasar-mode. If the Eddington ratio ($\dot{M}_{\textrm{BH}}/\dot{M}_{\textrm{edd}}$) exceeds a threshold ($M_{\textrm{BH}}$-dependent, but $\sim0.001$--$0.1$), the BH is in the quasar mode and injects a large amount of thermal energy into its surroundings.

Second, is the star formation (SF) model, specifically how the SFR of a gas cell is set. Gas above the threshold density $\rho_{\textrm{th}}$ is given a SFR of $m_{\textrm{cell}}/t_{*}(\rho)$, where,
\begin{equation*}
t_{*}(\rho)=2.2\,\Gyr \left(\frac{\rho}{\rho_{\textrm{th}}}\right)^{-0.5}\textrm{.}
\end{equation*}
The threshold density is approximately $0.1\,\textrm{cm}^{-3}$. This model was originally conceived because it matched well the observed Kennicut-Schmidt relation \citep{Kennicutt1998,2003MNRAS.339..289S}. As will be discussed later, this relation is calibrated against normal star-forming galaxies, and so may underpredict the SFR at high gas densities.

Using the public catalog, we select a sample of subhalos at $z=1.5$ (snapshot 40) according to the following criteria: (1) the subhalo is central (i.e., the most massive subhalo within its halo), and (2) the subhalo's stellar mass is between $10^{10}$ and $10^{10.5}\,\Msun/\textrm{h}$. There were a total of 168 subhalos that met both criteria. The chosen mass range is broadly consistent with the expected mass of the Milky Way from abundance matching at this redshift \citep{2013ApJ...771L..35V}. We choose to make our selection of galaxies at $z=1.5$ instead of at lower redshift because we wish to capture the \textit{formation} of multimodal structure. Mergers at lower redshift contribute very little to the Milky Way's disk stars \citep[e.g.,][]{2016ARA&A..54..529B}, and would act as a contaminant in our sample.

We then visually inspected the abundance distribution in the \MgFe{}-\FeH{} plane for this sample of subhalos. Few subhalos display multimodal structure and, when present, is much weaker compared to that observed in the Milky Way. We then apply a post-processing step to the \MgFe{} of the stars, adding a value of,
\begin{equation*}
  0.1\times\left(t_{1.5}-t_{\textrm{form}}\right)\textrm{,}
\end{equation*}
to each star particle, where $t_{1.5}$ is the age of the universe at $z=1.5$ ($\sim4.3\,\Gyr$), and $t_{\textrm{form}}$ is the formation time of the star particle. After this adjustment, which we justify in Section~\ref{ssec:sfe} based on TNG underpredicting the SFE of dense gas, we observe that the abundance structure becomes significantly more pronounced.

We show the abundance distributions of 16 random subhalos from our Milky Way-progenitor mass catalog, and the subhalo we selected, at $z=0$ in Appendix~\ref{app:rand_fig1}. Of these subhalos, some structure is present, but none have a strong bimodality as seen in the Milky Way. After the $\alpha$-enhancement, six of the additional galaxies shown seem to have bimodal features: 143882, 167392, 348901, 425719, 439099, and 465255. However, none are as prominent as the fiducial galaxy (392276). It is worth noting that the post-processing is not in itself responsible for the emergence of bimodality, as it only does so for a fraction of the galaxies shown in Appendix~\ref{app:rand_fig1}. Furthermore, if a galaxy had \alphaFe{} slowly and continuously decrease with time, then our $\alpha$-enhancement would not give rise to a bimodality since it is also a continuous adjustment. Appendix~\ref{app:rand_fig1} also shows the effect of choosing a different redshift at which to apply the enhancement ($z=1$, 1.5, and 2), which does not qualitatively change the bimodal structure in our fiducial galaxy.

We used \texttt{FOF} and \texttt{SUBFIND} in order to identify substructure \citep{2005Natur.435..629S,2009MNRAS.399..497D}. We select subhalo 172175 (the \texttt{SUBFIND} ID at snapshot 40) for its resemblance to the Milky Way. We then studied the main descendant of this subhalo at $z=0$ (subhalo 392276 at snapshot 99). A summary of its key properties is given in Table~\ref{tab:summ}.

% Please add the following required packages to your document preamble:
\begin{deluxetable}{lc}
  \tablecaption{Summary statistics of our TNG galaxy at $z=0$. $M_{200}$ and $R_{200}$ are relative to the mean density of the universe, the SFR is for all gas bound to the subhalo, and the BH refers to the central BH. \label{tab:summ}}
  \tablewidth{0pt}
  \tablehead{
    % \colhead{} & \colhead{}
  }
  \startdata
  \multicolumn{2}{c}{\textbf{dark matter}} \\ \hline
  $M_{200}$ & $4.97\times10^{12}\,\Msun$ \\
  $R_{200}$ & $532\,\kpc$ \\
  & \\
  \multicolumn{2}{c}{\textbf{baryons}} \\ \hline
  $\rhalf$ & $2\,\kpc$ \\
  $M_{\textrm{star}}(r<4\rhalf)$ & $7.6\times10^{12}\,\Msun$ \\
  $M_{\textrm{gas}}(r<4\rhalf)$ & $2.6\times10^{11}\,\Msun$ \\
  $\textrm{SFR}$ & $0.56\,\Msunyr$ \\
  $M_{\textrm{BH}}$ & $2.9\times10^{8}\,\Msun$ \\
  $\dot{M}_{\textrm{BH}}/\dot{M}_{\textrm{Edd}}$ & $4.5\times10^{-5}$ \\
  \enddata
\end{deluxetable}

\subsection{Observational Data}\label{ssec:obs}
We make use of two observational data sets. First, we use the stellar abundances from the ASPCAP pipeline of APOGEE DR17 \citep[][J.A.~Holtzman et al., in preparation]{2016AJ....151..144G,2017AJ....154...28B,2017AJ....154...94M,2022ApJS..259...35A}. We make the same selection cuts as in Section~2.4 of Paper I. These criteria select giants with high quality abundance measurements and angular momenta similar to the Sun's. This results in a sample of 54,777 stars. We use Fe to track total metallicity and Mg alone as an $\alpha$-element.

We then further considered a dataset of stellar ages from the APOKASC-3 catalog \citep{2024arXiv241000102P}. This catalog uses a combination of APOGEE spectroscopic parameters and \textit{Kepler} time series photometry to compute astroseismic ages. Using only stars with $25\%$ age uncertainties (taken as the maximum of the upper and lower uncertainty), we cross-match this catalog to our larger sample from ASPCAP which results in a sample of 2525 stars.

\subsection{One-Zone Chemical Evolution Model}\label{ssec:onezone_met}
As discussed in Section~\ref{ssec:tng}, we apply post-processing to enhance the $\alpha$-abundance of old star particles, motivated by the underprediction of the SFR in dense gas within TNG. To explore this, we use one-zone galactic chemical evolution (GCE) models, which describe enrichment in an idealized gas reservoir where newly synthesized metals mix instantaneously (see, e.g., the reviews by \citealt{Tinsley1980} and \citealt{Matteucci2021}). Our parameter choices in this paper are based on the models in \citet{2022arXiv220402989C}. While they were interested in bursts of star formation at the transition between the stellar halo and thick disk formation epochs, we focus on smooth evolutionary histories in this paper. We integrate these models numerically using the publicly available {\tt Versatile Integrator for Chemical Evolution} ({\tt VICE}; \citealt{2020MNRAS.498.1364J}).

The quantity of particular importance to our results is the depletion time $\tau_{\textrm{dep}}$, also known as the inverse of the star formation efficiency (SFE)\footnote{Note that the SFE we discuss here is averaged over a representative patch (e.g., kpc-size) and is not the dimensionless fraction of a giant molecular cloud that is converted into stars over its lifetime or local freefall time.} or the SFE timescale:
\begin{equation}
\tau_{\textrm{dep}} \equiv \frac{M_{\textrm{gas}}}{\SFR},
\end{equation}
where $M_{\textrm{gas}}$ is the mass of the gas reservoir. This timescale can be equivalently expressed as the ratio of the corresponding surface densities $\Sigma_{\textrm{gas}}$ and $\dot{\Sigma}_{\textrm{star}}$. In principle, $\tau_{\textrm{dep}}$ should vary with $M_{\textrm{gas}}$ because the observed relation between $\dot{\Sigma}_{\textrm{star}}$ and $\Sigma_{\textrm{gas}}$ is non-linear; classically, $\dot{\Sigma}_{\textrm{star}} \propto \Sigma_{\textrm{gas}}^N$ where $N \approx 1.5$ (e.g. \citealt{Schmidt1959, Kennicutt1998}). For simplicity, we hold $\tau_{\textrm{dep}}$ constant in this paper, which corresponds to a linear relation. In Section~\ref{ssec:onezone} below, we present models using $\tau_{\textrm{dep}} = 0.5$, $1$, and $5\,\Gyr$.

Following \citet{2022arXiv220402989C}, our models assume that there is initially no gas present in the galaxy (i.e. $M_{\textrm{gas}} = 0$ at $t = 0$). Zero metallicity gas accretes at a constant rate of $\dot{M}_\text{in} = 5\,\Msunyr$. The efficiency with which feedback sweeps up interstellar material and ejects it to the circumgalactic medium in an outflow is described by the mass loading factor $\eta \equiv \dot{M}_\text{out} / \SFR$. We use $\eta = 2$ throughout this paper (i.e. for every solar mass of stars formed, $2\,\Msun$ of interstellar gas is lost to an outflow). These parameter choices lead to a SFH that initially rises from zero and approaches a constant value of $\sim1.9\,\Msunyr$. With these choices, the predicted abundance evolution of individual elements is insensitive to the overall normalization of the accretion and star formation histories, since a larger accretion rate forms more stars and metals but simultaneously introduces more H into the interstellar medium. The evolution of \MgFe{} is unchanged by these parameters.

We use the solar abundances of Fe and Mg from \citet{Magg2022} with an additional $0.04\,\dex$ to account for the effects of diffusion and gravitational settling (\citealt{Turcotte1998}; i.e. $\{Z_{\text{Fe},\odot}, Z_{\text{Mg},\odot}\} = \{13.7, 6.71\} \times 10^{-4}$). We adopt SN yields of Fe and Mg based on \citeauthor{2024ApJ...973..122W}'s \citeyearpar{2024ApJ...973..122W} recommendations for Type~II SNe:
\begin{itemize}

	\item $y_\text{Fe}^\text{II} = 4.73 \times 10^{-4}$

	\item $y_\text{Mg}^\text{II} = 6.52 \times 10^{-4}$

	\item $y_\text{Fe}^\text{Ia} = 12 \times 10^{-4}$

	\item $y_\text{Mg}^\text{Ia} = 0$,

\end{itemize}
where the subscripts and superscripts denote the element and SN type, respectively. We select this value of $y_\text{Fe}^\text{Ia}$ as it corresponds to $\sim30\%$ ($\sim70\%$) of Fe arising from Type~II (Ia) SNe at solar \FeH{}. These yields are population-averaged, meaning that they quantify the amount of metal production per unit mass of star formation under the given enrichment channel (e.g. a hypothetical $10^3\,\Msun$ stellar population would produce a total of $0.652\,\Msun$ of Mg through Type~II SNe). {\tt VICE} models Type~II SNe as exploding instantaneously following an episode of star formation. For Type~Ia SNe, we assume that metals are produced according to the empirically motivated $t^{-1.1}$ power-law based volumetric SN rates as a function of redshift and the cosmic SFH \citep[e.g.][]{Maoz2012}.

We use the yields recommended by \citet{2024ApJ...973..122W} because they are based on an empirical determination of the mean $^{56}$Ni yield from Type~II SNe by \citet{Rodriguez2021, Rodriguez2023} applied to multi-element abundance distributions in the Milky Way. Theoretically predicted SN yields \citep[e.g.][]{Seitenzahl2013, Sukhbold2016, Limongi2018, Gronow2021} are subject to substantial theoretical uncertainties. Empirical considerations regarding yields are often necessary for GCE models to make tenable predictions (see also discussion in \citealt{Palla2022}).

With the exception of our SN yields, we find through experimentation that the decline in \MgFe{} with time is most significantly sensitive to the value of $\tau_{\textrm{dep}}$, as mentioned before. We therefore expect its connection to the decline in \MgFe{} on which this paper is focused to be robust. We refer to \citet{2020MNRAS.498.1364J} and \citet{2022arXiv220402989C} for further discussion of these GCE models.

\section{Results}\label{sec:results}
\subsection{Abundance Plane}\label{ssec:plane}

\begin{figure*}
  \centering
  \includegraphics[width=\textwidth]{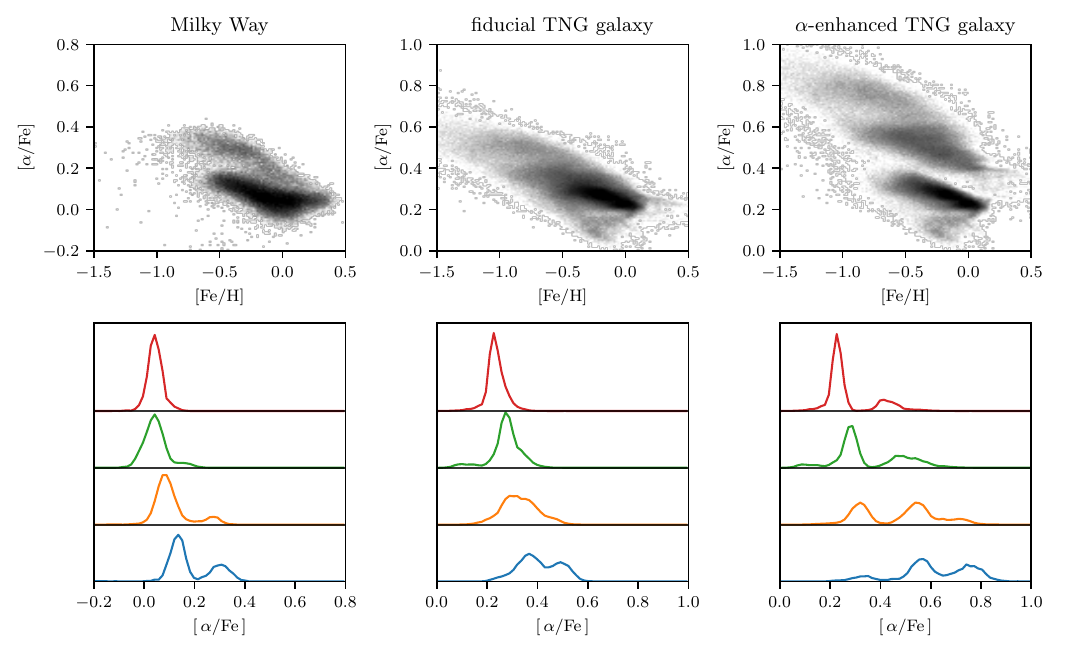}
  \caption{\textbf{When old stars are $\alpha$-enhanced, our galaxy from TNG displays a prominent bimodality.} The upper left panel shows the distribution in the \MgFe{}-\FeH{} plane of the Milky Way, demonstrating a clear bimodality (data selection given in text). The lower left panel shows the 1D histograms of \MgFe{} at fixed \FeH{} values of $-0.5$, $-0.25$, $0$, and $0.25$ (blue, orange, green, and red, respectively). In the Milky Way, the bimodality is strongest at low metallicities while disappearing at high metallicities. The middle column shows the same plots but for our TNG galaxy (392276) and with the fixed \FeH{} values $0.25\,\dex$ lower. Only faint structure is seen in the lowest bin (blue, $-0.75\,\dex$). The right column shows the same subhalo but after increasing the \MgFe{} value of star particles formed before $z=1.5$ linearly with formation time (with a slope of $0.1\dex/\Gyr$). A clear bimodality is shown in these panels which, unlike in the Milky Way, is present at all metallicities.}
  \label{fig:fig1}
\end{figure*}

The main result of our paper is given in Figure~\ref{fig:fig1}. Here, we compare the abundance plane in the Milky Way (left column) to that of our TNG galaxy (middle and right columns). The upper panels show the 2D distribution in the space of \MgFe{}-\FeH{}. We have applied the standard \texttt{scipy} implementation of a Gaussian kernel density estimator to a Cartesian grid of points. For each panel, we normalize so that the integral of the distribution is unity. Colors are plotted in a log scale ranging from $0.08$ to $15\,\dex^{-2}$. Dashed contour lines are plotted at $0.1$, $1.5$, and $10\,\dex^{-2}$.

The colored vertical regions are indicated at $\FeH=-0.75$, $-0.5$, $-0.25$, and $0\,\dex$ in the Milky Way, and at bins $0.25\,\dex$ higher in the simulations. The lower panels show 1D histograms of \MgFe{} in bins centered on these values. The bins have width $0.1\,\dex$, which is reflected in the width of the colored, shaded regions in the upper panels. The rationale for the higher plotted \FeH{} in the simulations reflects the empirical location of the bimodalities. The Milky Way shows a clear bimodal population, with a high-$\alpha$ sequence most clearly distinct from the low-$\alpha$ sequence at low metallicity. The two sequences merge around solar metallicity.

Our galaxy, on the other hand, does not show a clearly bimodal structure in the fiducial simulation (middle column). There is some structure in the $\FeH=-0.75$ bin. The right panel of Figure~\ref{fig:fig1} shows the same distribution as in the middle panel, but with a modification to increase \MgFe{} values of older stars formed before $z=1.5$ (see Sections~\ref{ssec:tng} and \ref{ssec:onezone}, as well as the upper panels of Figure~\ref{fig:alpha} for a visual demonstration). A multimodal structure emerges with three clear modes at $\MgFe\sim0.8$, $0.5$, and $0.2\,\dex$. The 1D histograms show that the modes are well-separated, and that the troughs between the modes nearly vanish.

\subsection{Alpha Time Dependence}\label{ssec:alpha_time}

\begin{figure*}
  \centering
  \includegraphics[width=\textwidth]{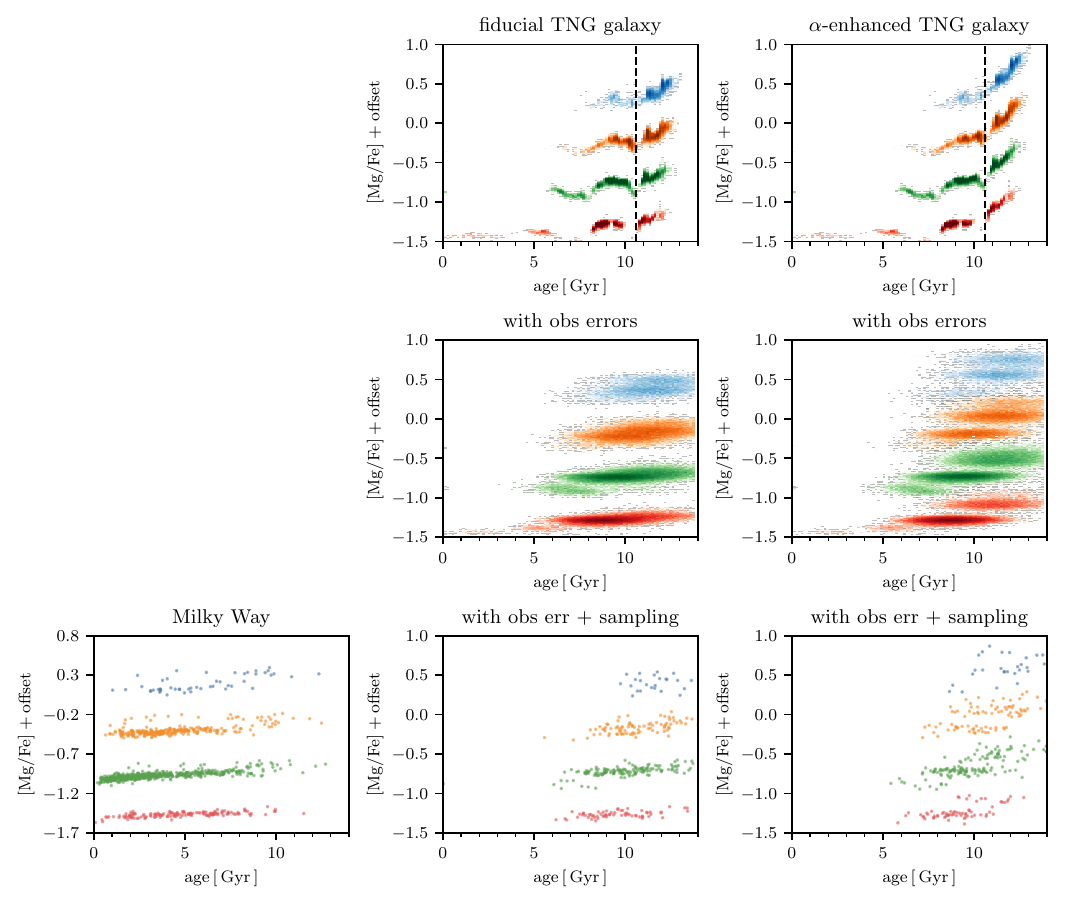}
  \caption{\textbf{Bimodality in the abundance plane is linked to distinct epochs separated by quiescence in simulation.} The upper row shows \MgFe{} as a function of age for our subhalo in TNG. The colors indicate stellar populations at fixed values of \FeH{}, which are the same as in Figure~\ref{fig:fig1}. A gap in the relation occurs at an age of approximately $10.6\,\Gyr$, which we indicate with a vertical dashed line. The effect of the $\alpha$-enhancement is clear, as it separates the stars that form before and after this gap in ages (star particles which formed before $z=1.5$ are $\alpha$-enhanced, which occurs at an age of $\sim9.5\,\Gyr$). The middle row shows the same TNG and $\alpha$-enhanced TNG data, but with added uncertainties of $12.5\%$ in age and $0.015\,\dex$ in \MgFe{}. When given these errors, the before and after star particles smear such that the two populations significantly overlap in ages. There is a second population of stars linked to another gap at $\sim8\,\Gyr$, discussed in the text. The lower row shows on the left the Milky Way data and on the right the same TNG data with artificial errors but subsampled to the same number of stars older than $5\,\Gyr$ as in the observations (because the simulation sample has almost no star particles younger than $5\,\Gyr$). The limited sample size of the observations makes a direct comparison difficult.}
  \label{fig:alpha}
\end{figure*}

The abundance distributions shown in Figure~\ref{fig:fig1} can be better understood by examining the evolution of \MgFe{} with time of the individual stars/star particles. In the upper panels of Figure~\ref{fig:alpha} we show the true distribution of \MgFe{} as a function of time for the fiducial galaxy in the middle and for the post-processed, $\alpha$-enhanced subhalo to the right. We use age instead of formation time in order to better facilitate comparisons to observations. These panels show 2D histograms, with a logarithmic colormap normalized to the maximum of the plot. To prevent overlap, the values of \MgFe{} are given offsets of $0$, $-0.5$, $-1$, and $-1.5\,\dex$, in order of increasing \FeH{}.

In our simulated galaxy, there is an age gap at $\sim10.6\,\Gyr$, which we mark with a vertical dashed line in the upper row. Star particles older than this line have a much clearer gradient in time with \MgFe{} than stars that form after, even in the fiducial case. In the $\FeH=-0.25$ bin, star particles which form directly after this line have a slightly reduced \MgFe{} than stars which form a short time later.

In the middle row, the center and right panels show the simulated galaxies with Gaussian errors of $12.5\%$ in age and $0.015\,\dex$ in \MgFe{}, aligning with the observational uncertainties in the APOKASC-3 and APOGEE datasets (see Appendix~\ref{app:obs_err}). The error in \MgFe{} is insignificant, but the age error ($1.25\,\Gyr$ at $10\,\Gyr$) significantly blurs the distribution, particularly across the dashed line that marks the transition between sequences. Despite this, the $\alpha$-enhanced galaxy still shows two distinct populations, although their ages now overlap more significantly.

The upper and middle row shows that there is a potential third population of star particles in the simulation, which is most visible in the $\FeH=-0.25$ and $0\,\dex$ bins (green and red, respectively). A minor gap in the upper panels is present at an age of $\sim8\,\Gyr$, which we discuss further in Section~\ref{ssec:evol}.

The lower row compares Milky Way data (left) with simulations of fiducial (middle) and $\alpha$-enhanced (right) galaxies. Here, the simulations have been subsampled to match the observed sample size of stars older than $5\,\Gyr$ in each metallicity bin. We match the sample size only to old stars because our simulated sample has almost no star particles younger than $5\,\Gyr$. When subsampled with observational errors, the $\FeH=-0.5$ bin in the simulation (orange right panel) very faintly shows a hint of two stacked distributions which might also be present in the $\FeH=-0.25$ bin in the data (orange left panel). The limited sample size makes it impossible to draw any strong conclusions.

\subsection{Evolutionary History}\label{ssec:evol}
\begin{figure}
  \centering
  \includegraphics[width=\columnwidth]{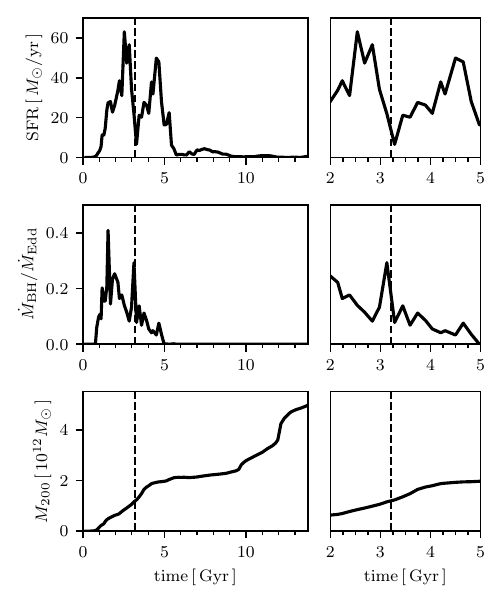}
  \caption{\textbf{The evolutionary history of our TNG galaxy.} The left column shows the SFH, BH accretion rate, and virial mass ($M_{200}$) over the entire time span, while the right column zooms in on the period from $t=2\,\Gyr$ to $t=5\,\Gyr$. The vertical dashed line in each panel marks the transition at $t\sim3.2\,\Gyr$, corresponding to the separation between the high- and low-$\alpha$ sequences (as shown in Figure~\ref{fig:alpha}). The upper panel shows the SFH, the middle panel shows the BH accretion rate as a fraction of the Eddington limit, and the lower panel shows the virial mass, representing the mass enclosed within a radius where the density is $200\times$ the mean cosmic density.}
  \label{fig:history}
\end{figure}

To understand the key events driving the behavior around the dashed line in Figure~\ref{fig:alpha}, we examine the evolution of several properties of our galaxy in Figure~\ref{fig:history}: its SFH, the BH accretion rate, and the growth of the virial mass. The vertical dashed line in each panel marks the transition at $t\sim3.2\,\Gyr$ from the high- to low-$\alpha$ sequences, as in Figure~\ref{fig:alpha}. The upper panel in the left column shows the the SFR (computed for all gas cells in the \texttt{SUBFIND} subhalo). There are two peaks at $t\sim2.5\,\Gyr$ and $t\sim4.5\,\Gyr$ with maximum values of $50\,\Msun/\textrm{yr}$ and $30\,\Msun/\textrm{yr}$, respectively. Around the high- to low-$\alpha$ transition, there is a dip in the SFR, which drops by an order of magnitude to about $3\,\Msun/\textrm{yr}$. The right panel zooms in on the period between $t=2\,\Gyr$ and $5\,\Gyr$, where we observe a sharp recovery in the SFR following the quiescent phase. In the span of a single snapshot (roughly $150\,\Myr$), the SFR increases from about $3\,\Msun/\textrm{yr}$ to $10\,\Msun/\textrm{yr}$.

There is also a more minor period of quiescence at a cosmic time of $\sim6\,\Gyr$, followed by a period of SF at a much lower rate. This is likely linked to the third population separated by an age gap at $\sim8\,\Gyr$ seen in Figure~\ref{fig:alpha}.

The middle panels track the accretion rate of the central BH as a fraction of the Eddington rate. Early in our galaxy's history ($t<2\,\Gyr$), the BH experiences high accretion, which steadily declines until $t\sim5\,\Gyr$. Around $t\sim3.2\,\Gyr$, the BH accretion rate peaks again, reaching approximately $30\%$ of the Eddington limit, placing the BH in quasar mode and injecting significant thermal energy into the galaxy's center. The middle right panel shows the period between $t=2$ and $5\,\Gyr$. We can see that the decline in the galaxy's SFR is contemporaneous with this increase in the BH accretion rate.

The lower panel illustrates the growth of the galaxy's virial mass ($M_{200}$). Early on ($t<4\,\Gyr$), $M_{200}$ increases roughly linearly, reaching about $2 \times 10^{12}\,\Msun$. After this, the mass remains relatively stable until jumps occur around $t\sim10$ and $\sim12\,\Gyr$, indicative of mergers. These late-time mergers raise the virial mass to $5 \times 10^{12},\Msun$, well above the typical Milky Way estimate of $1$--$1.5\times 10^{12},\Msun$ \citep[e.g.][]{2016ARA&A..54..529B}. However, during the high- to low-$\alpha$ transition, the virial mass remains consistent with a Milky Way progenitor, making this galaxy a suitable analog. There are no mergers related to the earlier quiescent period around $t\sim3.2\,\Gyr$ (no major mass jumps are observed during this time), implying the AGN activity is not merger-driven. The lower right panel shows the lack of mergers in more detail during the period between $t=2$ and $5\,\Gyr$.

\begin{figure}
  \centering
  \includegraphics[width=\columnwidth]{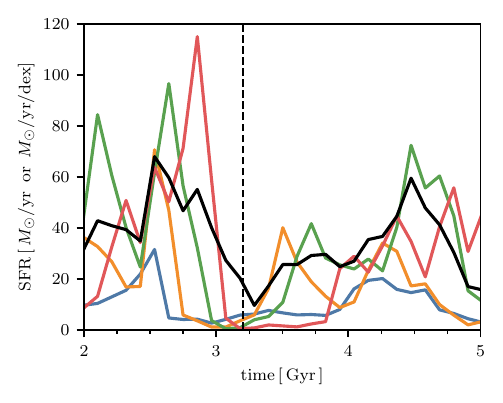}
  \caption{Global star formation history (black) compared with the star formation history in narrow metallicity bins, color-coded as in Figure~\ref{fig:fig1}. This plot demonstrates that the key condition for generating an $\alpha$-bimodality, the cessation of star formation in specific metallicity ranges, is satisfied. The metallicity-dependent SFR drops nearly to zero in every bin for periods extending up to $\sim1\,\Gyr$. The timing and duration of metallicity-dependent gaps can vary, preventing the total SFR from falling below $\sim10\,\Msunyr$.}
  \label{fig:zSFH}
\end{figure}

As discussed in Paper~I, the key to generating an $\alpha$-bimodality is halting star formation within specific metallicity ranges. A global quiescent period is a sufficient but not necessary condition for the formation of such gaps. To illustrate this point, we present Figure~\ref{fig:zSFH}, which compares the global SFH in black with the SFH in narrow metallicity bins, color-coded to match the bins shown in Figure~\ref{fig:fig1}. This plot uses the $z=0$ distribution of star particle formation times, resulting in minor differences from the SFH shown in Figure~\ref{fig:history}. Notably, we observe a significant drop in the SFR within these metallicity bins, lasting up to $\sim1\,\Gyr$. However, because the timing of these metallicity-dependent gaps differs, the total SFR never falls below $\sim10\,\Msunyr$.

\subsection{Bar-Driven Quenching}\label{ssec:sequence_of_events}
\begin{figure*}
  \centering
  \includegraphics[width=\textwidth]{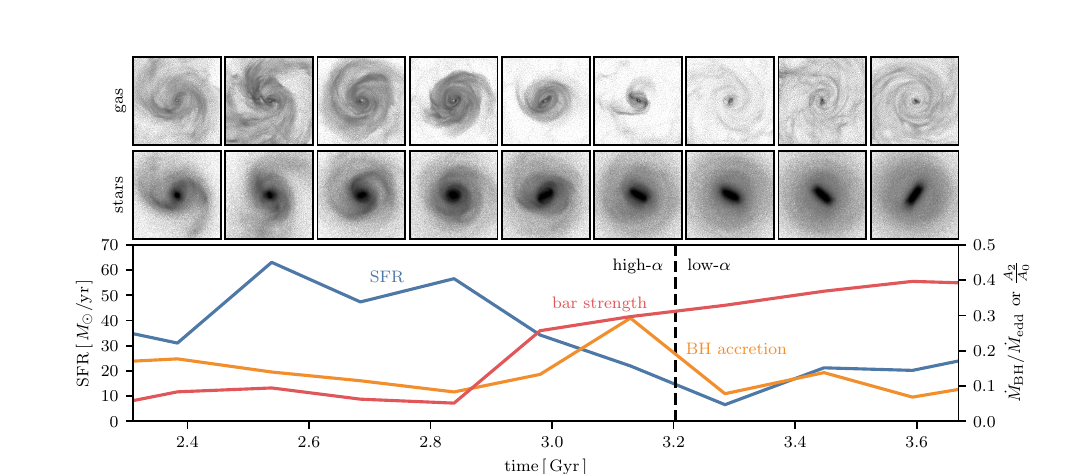}
  \caption{\textbf{Quiescence separating the high- and low-$\alpha$ sequences is preceded by AGN activity associated with bar formation.} Surface density projections of gas (top row) and star particles (middle row) in our galaxy across snapshots at different times during the high- to low-$\alpha$ transition. Below the projections is a plot showing the SFR, BH accretion rate (in units of $\dot{M}_{\textrm{BH}}/\dot{M}_{\textrm{edd}}$), and bar strength ($A_2/A_0$ for $R<2$ kpc). Time ranges from $\sim2.4\,\textrm{Gyr}$ to $\sim3.6\,\textrm{Gyr}$, corresponding to redshifts from $z\sim2.7$ to $z\sim1.8$. A sequence of events in which the bar strengthens, BH accretion increases, and SFR declines is seen, and is described more fully in the text.}
  \label{fig:seq}
\end{figure*}

We find that a quenching episode is driven by the formation of a bar contemporaneous with an increase in AGN activity, shown in Figure~\ref{fig:seq}. The upper panels show gas density, while the middle panels display stellar density. Time progresses from $\sim2.4$ to $\sim3.6\,\textrm{Gyr}$, corresponding to redshifts ranging from $z\sim2.7$ to $z\sim1.8$, and the high- to low-$\alpha$ transition is indicated with a vertical dashed line.

This figure shows the following sequence of events:
\begin{enumerate}
    \item Bar forms: A steady increase in the bar strength, as indicated by $A_2/A_0$ for star particles with $R<2\,\kpc$, from $\sim0.05$ to $0.4$ starting around $2.8\,\textrm{Gyr}$. This rise is accompanied by the appearance of elongated features in the gas and stars consistent with a bar.
    \item BH accretion increases: Following the increase in bar strength by about a snapshot ($\sim150\,\Myr$ here), the BH accretion rate ($\dot{M}_{\textrm{BH}}/\dot{M}_{\textrm{edd}}$) shows a significant spike, rising from a minimum of $\sim0.08$ at $t=2.84\,\Gyr$ to a maximum of $\sim0.29$ at $t=3.13\,\Gyr$ for one snapshot.
    \item SFR declines: The SFR declines starting from a maximum of $54.3\,\Msunyr$ at $t=2.84\,\Gyr$ down to $3.6\,\Msunyr$ at $t=3.28\,\Gyr$. In the next snapshot at $t=3.45\,\Gyr$ the SFR recovers to $12.2\,\Msunyr$. Figure~\ref{fig:history} shows that it reaches its second maximum of $30.9\,\Msunyr$ at $4.5\,\Gyr$. We show in Figure~\ref{fig:zSFH} that the global SF suppression is coincident with a much more dramatic metallicity dependent SF gap, with the SFR dropping nearly to $0\,\Msunyrdex$ in several bins.
\end{enumerate}

Note that in the Milky Way the bar is estimated to have formed approximately $8\,\Gyr$ ago \citep{2019MNRAS.490.4740B,2024MNRAS.530.2972S}, coinciding with the epoch when the bimodality is observed to emerge \citep{2013A&A...560A.109H,2023MNRAS.525.2208R,2024MNRAS.535..392L}.

\subsection{\alphaFe{} with Varying SFE}\label{ssec:onezone}

\begin{figure}
  \centering
  \includegraphics[width=\columnwidth]{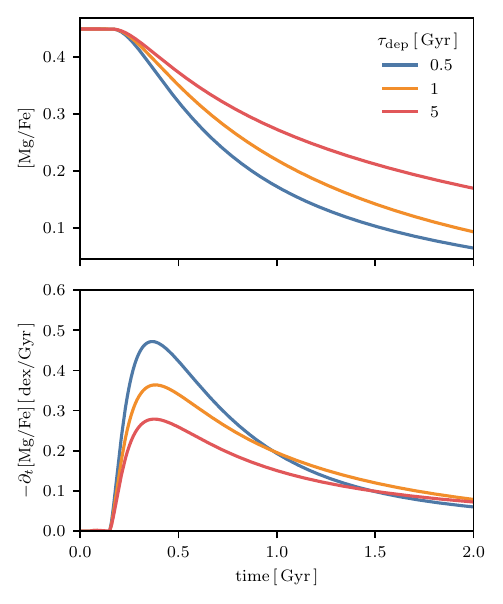}
  \caption{\textbf{A higher star formation efficiency leads to a steeper decline in \MgFe{}.} In both panels, the lines show the time evolution of \MgFe{} in a simple one-zone chemical evolution model, described in Section~\ref{ssec:onezone_met}. The evolution for different SFE values of $2$, $1$, and $0.2\,\textrm{Gyr}^{-1}$ are shown in blue, orange, and red, respectively.  The upper panel shows the evolution of \MgFe{} over $2\,\Gyr$, while the lower panel shows the negative of its time derivative. Increasing the SFE (decreasing $\tau_{\textrm{dep}}$) leads to a more rapid decline in \MgFe{}. At its steepest decline ($t\sim0.5\,\Gyr$), increasing the SFE by an order of magnitude results in a slope nearly twice as steep. At later times ($t > 1\,\Gyr$), models with higher SFE reach their steady-state \MgFe{} value more quickly.}
  \label{fig:vice}
\end{figure}

In Figure~\ref{fig:fig1}, we showed that a time-linear $\alpha$-enhancement of older stars (formed before $z=1.5$) leads to the emergence of a pronounced chemical bimodality. This $\alpha$-enhancement corresponds to a more rapid decline in \alphaFe{} over time at high redshifts. Here we show that the steeper \alphaFe{} evolution implied by this $\alpha$-enhancement of old stars can be physically justified as a boost to the SFE of dense gas.

The evolution of \MgFe{} in three one-zone GCE models with varying SFEs is shown in the upper panel of Figure~\ref{fig:vice} (models described in Section~\ref{ssec:onezone_met}). A higher SFE (lower $\tau_{\textrm{dep}}$) leads to a faster reduction in \MgFe{}. In the model with the highest SFE ($2\,\Gyr^{-1}$, $\tau_{\textrm{dep}}=0.5\,\Gyr$, blue), \MgFe{} drops from $\sim0.45$ to $0.08\,\dex$ over $2\,\Gyr$. In contrast, the model with the lowest SFE ($0.2\,\Gyr^{-1}$, $\tau_{\textrm{dep}}=5\,\Gyr$, red) only reaches $\sim0.2\,\dex$ within the same period.

The lower panel shows the negative time derivative of \MgFe{} (i.e., the rate of decline). The model with the highest SFE ($2\,\Gyr^{-1}$, blue) has a peak decline rate of $-0.5\,\dex/\Gyr$, while the model with the lowest SFE ($0.2\,\Gyr^{-1}$) peaks at $-0.25\,\dex/\Gyr$. After $1\,\Gyr$, the trend begins to reverse, and the lower-SFE models catch up, though at a much slower rate ($\sim-0.1\,\dex/\Gyr$) compared to their peak.

This analysis illustrates that a higher SFE at early times (high-$z$) leads to a faster decline in \MgFe{}. Recent work has suggested that the SFE in such dense regions in TNG may indeed be too low, as discussed in Section~\ref{ssec:sfe}. The post-processing $\alpha$-enhancement in Section~\ref{ssec:tng} is meant to mimic a SFE correction of high-$z$ dense gas.

\section{Discussion}\label{sec:disc}
In Figure~\ref{fig:fig1}, we compared the abundance plane between the Milky Way and our galaxy before and after our $\alpha$-enhancement post-processing procedure. It is clear that the TNG galaxy is unimodal before the $\alpha$-enhancement and multi-modal afterwards. Here, we briefly discuss two main points: (1) assuming the $\alpha$-enhancement is justified, what leads to the bimodality in the TNG galaxy?, and (2) what justifies the $\alpha$-enhancement? We then extend our comparison to data and argue the quiescent period is driven by a bar-induced AGN episode.

\subsection{Cause of Bimodality}\label{ssec:bim_cause}
We argue that the available evidence for the cause of the bimodality aligns with the scenario outlined in Paper~I. That study presented an idealized simulation resembling the merger between the Milky Way and GSE, where the orbital parameters were varied slightly across a grid of 27 simulations. The results showed that simulations featuring a brief quiescent period produced a bimodal abundance distribution. Two conditions for the bimodality to arise have to be met: (1) a declining \alphaFe{} with time (here done through our post-processing step, see Figure~\ref{fig:alpha}), and (2) a gap in the metallicity-dependent SFR. In Paper~I, the gap in SFR was due to a merger, but here we argue it is due to the formation of a bar.

The galaxy that we have studied in this work is consistent with the quiescent scenario proposed in Paper~I. The distribution of \MgFe{} with star particle age is a useful way to test this scenario, which we plot in the upper right panel of Figure~\ref{fig:alpha} in fixed bins of \FeH{} (each color is a different \FeH{} bin). The vertical dashed line at $z\sim2$ in Figures~\ref{fig:alpha} and \ref{fig:history} marks the transition between the high- and low-$\alpha$ sequences.
\footnote{The transition between the high- and low-$\alpha$ sequences occurs approximately $1\,\Gyr$ before $z=1.5$, where the $\alpha$-enhancement begins. This, combined with the fact that not all of the subhalos in our sample display bimodalities (Appendix~\ref{app:rand_fig1}), shows that the $\alpha$-enhancement is not the direct cause of the bimodality.} At the transition, there is a $\sim300\,\Myr$ quiescent period during which the SFR drops by a factor of $\sim10$ to $15$.

This global quiescent period is symptomatic of a more dramatic reduction in the star formation rate in narrow metallicity bins, which almost entirely vanishes (Figure~\ref{fig:zSFH}). As we showed in Paper~I, it is the metallicity-dependent quiescence in the presence of a declining gas-phase \MgFe{} -- not global quiescence -- that leads to an $\alpha$-bimodality. In the fiducial TNG model, the \MgFe{} reduction with time is not rapid enough, leading to the post-processing discussed in Section~\ref{ssec:sfe}. In the real universe, a global quiescent period of a duration of $\sim300\,\Myr$ would lead to a reduction in enrichment from Type~II relative to Type~Ia SNe leads to a lower rate of Mg production. The typical lifetime of a Type~II SN progenitor in this model is $\sim40\,\Myr$ \citep{2018MNRAS.473.4077P}, and so the hundreds of Myr of suppressed SF would be short enough to restrict the production of $\alpha$-elements.

In the fiducial TNG distribution, shown in the upper middle panel of Figure~\ref{fig:alpha}, the same general behavior is present. However, because the \MgFe{} decline before the quiescent period is slower, star particles which formed before and after this period overlap in the \MgFe{} distribution shown in Figure~\ref{fig:fig1}.

Notably, both the fiducial and $\alpha$-enhanced galaxy show a slight rebound effect in \MgFe{}. The star particles which form directly after the dashed line when the SFR has just recovered have a slightly lower \MgFe{} than stars which form later on, by about $0.1$ to $0.2\,\dex$. We argue that it is plausible that during the period of suppressed star formation, the \alphaFe{} ratio of the star-forming gas drops sharply due to the reduced contribution of Type~II relative to Type~Ia SNe. Later, the \alphaFe{} of the gas will recover when the SFR also recovers, but there is a brief window when old, low-$\alpha$ stars can form. A similar behavior was seen in the one-zone models with bursty SFHs in \citet{2020MNRAS.498.1364J}.

\subsection{Motivation for $\alpha$ Post-processing}\label{ssec:sfe}
As described in Section~\ref{ssec:tng}, we applied a post-processing which increased the \MgFe{} of star particles at $z>1.5$ in a time-linear manner, discussed in Section~\ref{ssec:tng}. The post-processed subhalo is presented alongside the fiducial subhalo in the right and middle columns, respectively, of Figures~\ref{fig:fig1} and \ref{fig:alpha}.

The \MgFe{} value of star forming gas is the result of a complicated mixture of many different aspects of the TNG model, to name a few: stellar and AGN feedback which alter gas inflows and outflows, secular, dynamical evolution, SF prescription, magnetic fields, (lack of) cosmic rays, diffusivity of hydrodynamics solver, and, of course, enrichment models. Isolating the cause of the potentially too shallow evolution of \MgFe{} vs. time at high-$z$ is not straightforward nor, in our opinion, even possible. However, we do offer one reasonable explanation: the SFE at high densities, more present at high-$z$, is too low.

We demonstrate the impact of the SFE on the \alphaFe{} ratio using a simple one-zone chemical evolution model with the publicly available code \texttt{VICE}. The details of our setup is given in Section~\ref{ssec:onezone_met}. We vary the SFE of the model ($\textrm{SFR}/M_{\textrm{gas}}$), and examine the impact on \MgFe{} as a function of time. We find that a higher SFE does lead to a more rapid reduction in \MgFe{}. The rate of decrease in \MgFe{}, at its maximum, varies between $\sim-0.25$, $-0.35$, and $-0.5\,\dex/\Gyr$ in the $\textrm{SFE}=0.2$, $1$, and $2\,\Gyr^{-1}$ models, respectively. For our post-processing, we assumed an additional decrease rate of $0.1\,\dex/\Gyr$. Such a difference is well within the range of \MgFe{} slopes seen in our different $\tau_{\textrm{dep}}$ models, implying a factor of only $\sim2$ to $5$ in the SFE is needed to reach our post-processing slope.

The TNG model predicts depletion times at high gas surface densities ($\tau_{\textrm{dep}}\sim0.5-1\,\Gyr$ at $\Sigma_{\textrm{gas}} ~ 100-300\,\Msun/\pc^2$) which are a factor of $\sim2$--$3$ longer than derived for starburst galaxies at similar densities of $\tau_{\textrm{dep}}\sim30$--$300\,Myr$ assuming \citet{2013ARA&A..51..207B} X(CO) \citep[see][]{2019ApJ...872...16D,2021ApJ...908...61K,2024arXiv240909121H}. This is well within our needed factor of $\sim2$ to $5$ in the SFE (see Figure~\ref{fig:vice}). Therefore, \MgFe{} should decline more rapidly with a different feedback model (or future iteration of the TNG model) that leads to a higher SFE at high densities.

An intuitive understanding of the impact the decline in \alphaFe{} vs. time has is that, when \alphaFe{} declines rapidly, it is a better estimator of age. When \alphaFe{} is a better estimator of age, events which are separated temporally become better separated in the abundance plane.

\subsection{Direct Comparison to Observations}\label{ssec:compare_obs}
We attempted a direct comparison between our TNG galaxy and the Milky Way in the lower row of Figure~\ref{fig:alpha} (see Section~\ref{ssec:alpha_time}). In the simulations, the presence of the bimodality is obvious from a stacked distribution. However, this distribution is not obvious in the observations, primarily because the sample size of observations at old ages where the bimodality is present ($>5\,\Gyr$) is quite small. Nonetheless, the presence of a clean gap in stellar ages between the high- and low-$\alpha$ is completely washed out by the age uncertainties.

There is also a population of young, $\alpha$-rich stars in the APOKASC-3 data. These may or may not reflect the typical or average ISM chemistry. Arguments have been made that they are old stars with misclassified astroseismic ages due to binary mass transfer \citep[and references therein]{2023A&A...671A..21J}. However, some appear to be genuinely young \citep[and references therein]{2024arXiv241002962L}, with a range of explanations given \citep[e.g.][]{2015A&A...576L..12C,2021MNRAS.508.4484J,2023arXiv231105815S}. Disentangling these effects is far from clear and beyond the scope of this work. At least some of the young, $\alpha$-rich stars in Figure~\ref{fig:alpha} do not reflect the ISM chemistry at their inferred astroseismic age, and so would not be included in the TNG model. With this caveat, the two appear to be consistent.

\subsection{Cause of Quiescence}\label{ssec:cause_qui}
In Paper~I, AGN activity from a merger was the suspected cause for the quiescent period. In our galaxy here, there is indeed a brief burst in AGN accretion at the time of the merger (middle panel of Figure~\ref{fig:history}). Based on this burst, it is also reasonable to suspect that AGN activity is also responsible for the quiescent period in our galaxy, noting that the real cause of the formation of the $\alpha$-bimodality is a metallicity-dependent quiescent period (see Paper~I and Figure~\ref{fig:zSFH}). However, we argued in Section~\ref{ssec:sequence_of_events} that the cause of the localized spike in the BH accretion rate is not due to a merger but instead due to the formation/strengthening of a bar. This connection is further supported by estimates that the Milky Way's bar formed $\sim8\,\Gyr$ ago \citep[e.g.,][]{2019MNRAS.490.4740B,2024MNRAS.530.2972S}, roughly concurrent with the formation of the bimodality \citep{2013A&A...560A.109H,2023MNRAS.525.2208R,2024MNRAS.535..392L}.

There is a significant body of theoretical and observational work in support of this picture. Bars and other non-axisymmetric features have long been argued to funnel gas into the centers of galaxies on theoretical grounds \citep{1989Natur.338...45S,2010MNRAS.407.1529H}. It was recently shown by \citet{2024arXiv240906783F} that bars can induce quiescence by accelerating the growth of a SMBH, but they found there can be many Gyr between bar formation and quenching. In observations, barred galaxies preferentially host AGN in star-forming galaxies \citep{2012ApJS..198....4O,2022A&A...661A.105S}.\footnote{\citet{2022A&A...668L...3L} studied a galaxy from TNG100 in which a merger induced AGN activity that ejected gas from its center. A bar then formed out of the gas-poor disk.} Furthermore, at high-$z$, the AGN mechanism is thought to be responsible for quenching \citep[e.g.][and references therein]{2023arXiv230806317D,2024arXiv240417945P,2024ApJ...968L..21M,2024Natur.630...54B}.

Since barred galaxies preferentially host AGN, we therefore predict that barred galaxies would preferentially host $\alpha$-bimodalities. The GECKOS survey, which aims to constrain the $\alpha$-bimodality of a sample of edge-on galaxies, about half of which are barred, using integral field spectroscopy at different altitudes, could test this \citep[and J. v.~d.~Sande, private communication]{2024IAUS..377...27V}. On the other hand, the strength of a bar is not associated with the strength of the host AGN \citep[e.g.]{2022A&A...661A.105S}. So, it is not clear that bimodalities would be associated with bar strength.

A complicating factor for this picture comes from the high SFR associated with the galaxy. The depletion time ($M_{\textrm{gas}}/\textrm{SFR}$) at the $t=2.84\,\Gyr$ snapshot is $204\,\Myr$, shorter than the time it takes for the SFR to reach its minimum. This implies the possibility of starvation as a quenching mechanism. However, the average depletion time in the preceding 5 snapshots (which are each $\sim150\,\Myr$ apart) is $220\,\Myr$, so clearly the galaxy is accreting high amounts of gas from its environment. A definitive account, difficult with the current simulation outputs because of its sparse snapshot spacing, requires further work. We also delay to future work the cause of the secondary quiescence period at $\sim6\,\Gyr$, briefly discussed in Section~\ref{ssec:evol}.

\section{Conclusions}\label{sec:conc}
In this work, we examined a galaxy in Illustris TNG50 which is at a Milky Way-progenitor mass at $z=1.5$. After applying a post-processing step that increased the \MgFe{} of star particles formed before $z=1.5$, this subhalo hosts a strong bimodality in the plane of \MgFe{} and \FeH{}, shown in Figure~\ref{fig:fig1}. This post-processing is justified by arguing that the SFE of dense gas is too low in TNG \citep[][see discussion in our Section~\ref{ssec:sfe}]{2024arXiv240909121H}.

The formation of the bimodality, when the galaxy transitions from producing high- to low-\MgFe{} star particles (Figure~\ref{fig:alpha}), is coincident with both a global and metallicity-dependent suppression of star formation (Figures~\ref{fig:history} and \ref{fig:zSFH}). This suppression of star formation is preceded by the formation of a bar and subsequent AGN activity (Figure~\ref{fig:seq}). This scenario is plausible for the Milky Way, as recent estimates indicate that the bar formed around $8\,\Gyr$ ago \citep{2019MNRAS.490.4740B,2024MNRAS.530.2972S}, coinciding with the onset of the bimodality.

The lack of star formation in a narrow metallicity bin in the presence of a decline in \MgFe{} naturally leads to a gap in the present-day distribution of \MgFe{} (see Paper~I). However, in the fiducial TNG  model, the decline in \MgFe{} is not rapid enough during the period of quiescence to produce an $\alpha$-bimodality. Our post-processing $\alpha$-enhancement artificially injects this declination. In the real universe, a global reduction in the SFR may expedite the drop in \MgFe{} since the number of Type~II relative to Type~Ia SNe would drop.

This work adds further support to a scenario in which a quiescent period in the Milky Way's past is a plausible explanation for the Milky Way's abundance bimodality. We argued in Paper~I that the GSE merger could trigger this period. In this work we have argued that the formation of the Milky Way's bar could be responsible. Regardless of this scenario's relevance to the Milky Way, we also predict that the presence of a bar in external galaxies is correlated with the presence of an $\alpha$-bimodality. Because of observational errors in age, it is not possible to make a direct comparison between the simulated galaxy and dataset (Figure~\ref{fig:alpha} and Section~\ref{ssec:compare_obs}). In the future, more numerous and precise age estimates of old, metal-poor stars may be able to distinguish the formation scenarios of the $\alpha$-bimodality.

\section*{acknowledgements}
  We would like to thank the referee for a thoughtful and helpful report. We would like to thank Marc Pinsonneault and Jesse van de Sande for helpful discussions, and M. Pinsonneault for sharing early access to the APOKASC-3 dataset. AB would like to thank Todd Phillips for helpful discussions.

  JWJ acknowledges support from a Carnegie Theoretical Astrophysics Center postdoctoral fellowship. Support for VS was provided by Harvard University through the Institute for Theory and Computation Fellowship. LH acknowledges support by the Simons Collaboration on ``Learning the Universe.''

  This work has made use of data from the European Space Agency (ESA) mission {\it Gaia} (\url{https://www.cosmos.esa.int/gaia}), processed by the {\it Gaia} Data Processing and Analysis Consortium (DPAC, \url{https://www.cosmos.esa.int/web/gaia/dpac/consortium}). Funding for the DPAC has been provided by national institutions, in particular the institutions participating in the {\it Gaia} Multilateral Agreement.
  
  Funding for the Sloan Digital Sky 
  Survey IV has been provided by the 
  Alfred P. Sloan Foundation, the U.S. 
  Department of Energy Office of 
  Science, and the Participating 
  Institutions. 
  
  SDSS-IV acknowledges support and 
  resources from the Center for High 
  Performance Computing  at the 
  University of Utah. The SDSS 
  website is www.sdss4.org.
  
  SDSS-IV is managed by the 
  Astrophysical Research Consortium 
  for the Participating Institutions 
  of the SDSS Collaboration including 
  the Brazilian Participation Group, 
  the Carnegie Institution for Science, 
  Carnegie Mellon University, Center for 
  Astrophysics | Harvard \& 
  Smithsonian, the Chilean Participation 
  Group, the French Participation Group, 
  Instituto de Astrof\'isica de 
  Canarias, The Johns Hopkins 
  University, Kavli Institute for the 
  Physics and Mathematics of the 
  Universe (IPMU) / University of 
  Tokyo, the Korean Participation Group, 
  Lawrence Berkeley National Laboratory, 
  Leibniz Institut f\"ur Astrophysik 
  Potsdam (AIP),  Max-Planck-Institut 
  f\"ur Astronomie (MPIA Heidelberg), 
  Max-Planck-Institut f\"ur 
  Astrophysik (MPA Garching), 
  Max-Planck-Institut f\"ur 
  Extraterrestrische Physik (MPE), 
  National Astronomical Observatories of 
  China, New Mexico State University, 
  New York University, University of 
  Notre Dame, Observat\'ario 
  Nacional / MCTI, The Ohio State 
  University, Pennsylvania State 
  University, Shanghai 
  Astronomical Observatory, United 
  Kingdom Participation Group, 
  Universidad Nacional Aut\'onoma 
  de M\'exico, University of Arizona, 
  University of Colorado Boulder, 
  University of Oxford, University of 
  Portsmouth, University of Utah, 
  University of Virginia, University 
  of Washington, University of 
  Wisconsin, Vanderbilt University, 
  and Yale University.

We acknowledge the use of OpenAI’s ChatGPT and Anthropic's Claude for assistance in editing this manuscript for clarity and conciseness and in generating small analysis scripts and code snippets.

% \end{acknowledgements}

\bibliography{ref}{}
\bibliographystyle{aasjournal}

\appendix

\section{Observational Errors}\label{app:obs_err}
\begin{figure*}
  \centering
  \includegraphics[width=\columnwidth]{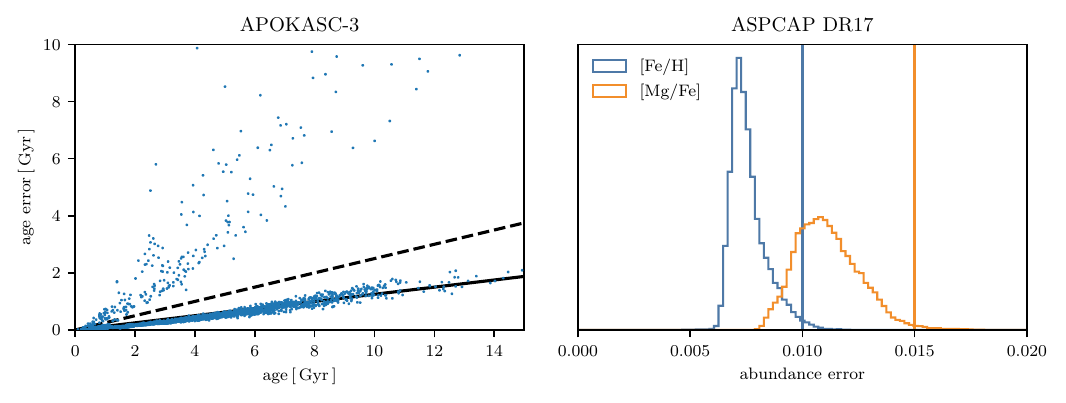}
  \caption{The observational errors of astroseismic ages from APOKASC-3 (left) and abundances from ASPCAP DR17 (right). We show, on the left, a line indicating a $12.5\%$ error in observed age and on the right a vertical line indicating a $0.01$ and $0.015\,\dex$ error in \FeH{} and \MgFe{}, respectively. On the left, a dashed line indicates the $25\%$ error cut used for inclusion in Figure~\ref{fig:alpha}.}
  \label{fig:obs_err}
\end{figure*}

In Figure~\ref{fig:alpha}, we assumed observational errors of $12.5\%$ in age and $0.015\,\dex$ in \MgFe{}. In Figure~\ref{fig:obs_err}, we plot the quoted observational errors of the APOKASC-3 (left) and ASPCAP (right) datasets, showing both \FeH{} and \MgFe{} (blue and orange, respectively). We show our $12.5\%$ age error as a black line in the left panel. We take the age error to be the maximum of the upper and lower errors from \citet{2018ApJS..239...32P}. In the right panel, we show blue and orange vertical lines at $0.01$ and $0.015\,\dex$ for \FeH{} and \MgFe{}, respectively. These are approximately the 99th percentiles of each error distribution. As a dashed line we show our $25\%$ age error cut for stars plotted in Figure~\ref{fig:obs_err}. Our assumed errors are generally consistent with the errors quoted in the dataset, with a conservative estimate for the abundance errors.

\section{Random Selection of Subhalos}\label{app:rand_fig1}
\begin{figure*}
  \centering
  \includegraphics[width=\textwidth]{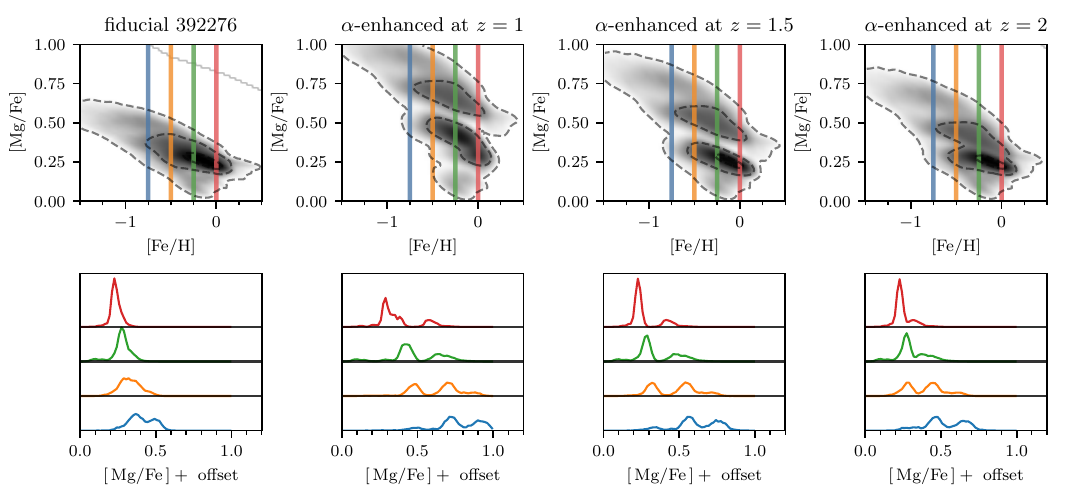}
  \caption{Abundance plane of the fiducial galaxy at $z=0$ and the effects of $\alpha$-enhancement applied at different redshifts. The leftmost panel shows the fiducial galaxy without enhancement, reproducing the middle column of Figure~\ref{fig:fig1}. The subsequent panels from left to right show the results of applying $\alpha$-enhancement at redshifts of 1, 1.5, and 2, respectively. The $z=1.5$ column reproduces the right column of Figure~\ref{fig:fig1}. The presence of substructure in the abundance plane is consistent across different application times of $\alpha$-enhancement, though applying it at lower redshifts appears to slightly increase the separation between modal peaks. This figure is part of a set of 17 images available in the online journal, showing similar plots for 16 additional galaxies from our Milky Way-progenitor mass sample. The varied responses to $\alpha$-enhancement across the sample, with only six additional galaxies (143882, 167392, 348901, 425719, 439099, and 465255) displaying bimodalities, suggest that while $\alpha$-enhancement generally increases substructure, it is not solely responsible for creating bimodalities in abundance planes.}
  \label{fig:app0}
\end{figure*}

In Figure~\ref{fig:app0}, we show the abundance plane of our fiducial galaxy at $z=0$. This reproduces the middle column of Figure~\ref{fig:fig1}. We also show the effect of our $\alpha$-enhancement procedure on this distribution when applied, from left to right, at redshifts of $1$, $1.5$, and $2$. (The $z=1.5$ column reproduces the right column of Figure~\ref{fig:fig1}). Qualitatively, the time at which the $\alpha$-enhancement is applied does not alter whether substructure arises in this plane. However, when it is applied at lower $z$, the peaks between modes do appear to be slightly further apart.

We show the same figure but with an additional random selection of 16 galaxies in the Milky Way-progenitor mass sample as a figure set (17 images), which is available in the online journal. Six additional galaxies (143882, 167392, 348901, 425719, 439099, and 465255) display bimodalities, though none as prominent as the main galaxy studied in this work. Some substructure is present in many galaxies. In general, the $\alpha$-enhancement increases the strength of substructure in the abundance planes. The fact that bimodalities like in the main galaxy studied in this work do not universally appear in $\alpha$-enhanced galaxies indicates that the $\alpha$-enhancement is not solely responsible for the bimodality.

The timing of the $\alpha$-enhancement does not have a major effect on our fiducial galaxy (Figure~\ref{fig:app0}). However, for some (e.g., the green $\FeH=-0.25$ bin in 439099), structure arises only when the $\alpha$-enhancement is applied at sufficiently low redshift. We interpret this as the presence of some substructure inducing activity between $z=1$ and 2.

\figsetstart
\figsetnum{8}
\figsettitle{Abundance plane for a random subset of galaxies in our Milky Way-progenitor mass sample, with and without $\alpha$-enhancement.}
\figsetgrpstart
\figsetgrpnum{8.1}
\figsetgrptitle{Subhalo 392276.}
\figsetplot{app_392276.pdf}
\figsetgrpnote{Abundance plane for subhalo 392276.}
\figsetgrpend
\figsetgrpstart
\figsetgrpnum{8.2}
\figsetgrptitle{Subhalo 2.}
\figsetplot{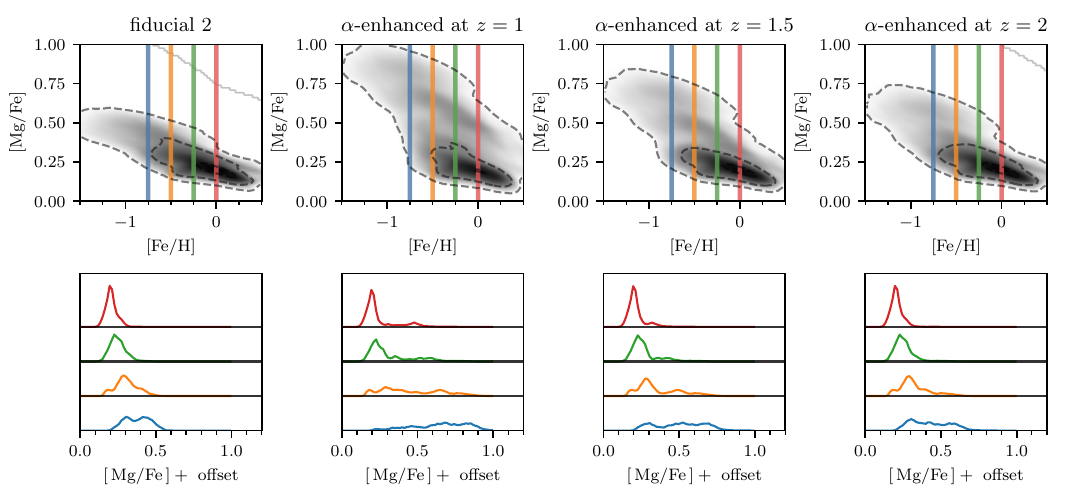}
\figsetgrpnote{Abundance plane for subhalo 2.}
\figsetgrpend
\figsetgrpstart
\figsetgrpnum{8.3}
\figsetgrptitle{Subhalo 10.}
\figsetplot{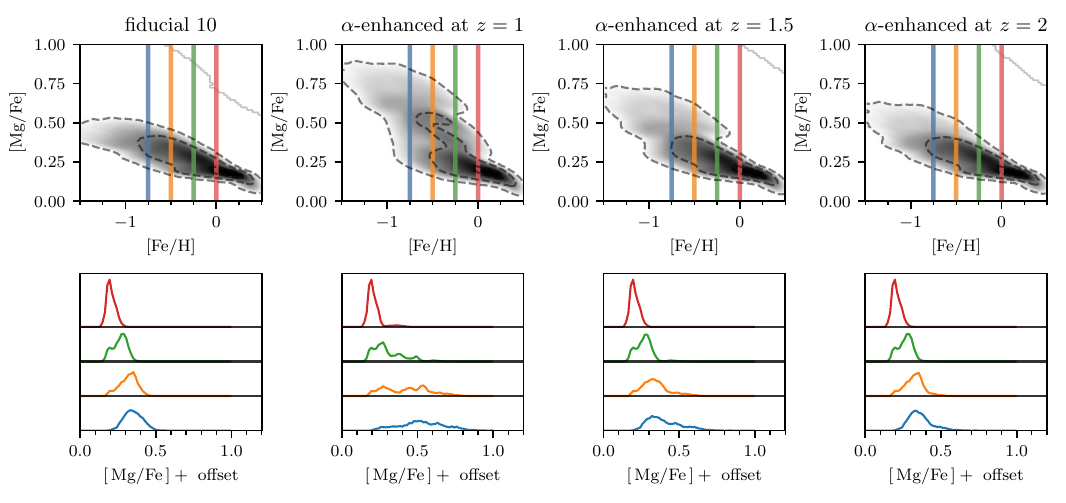}
\figsetgrpnote{Abundance plane for subhalo 10.}
\figsetgrpend
\figsetgrpstart
\figsetgrpnum{8.4}
\figsetgrptitle{Subhalo 143882.}
\figsetplot{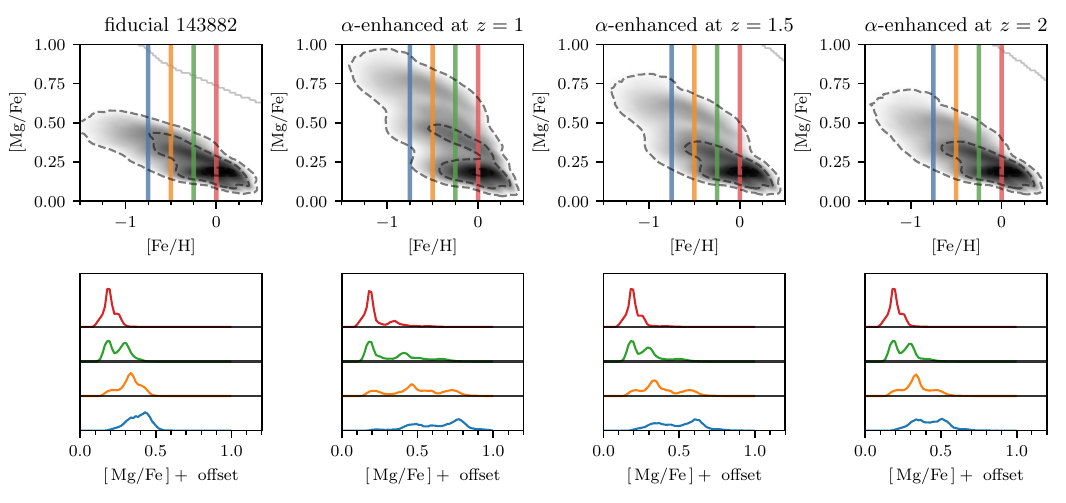}
\figsetgrpnote{Abundance plane for subhalo 143882.}
\figsetgrpend
\figsetgrpstart
\figsetgrpnum{8.5}
\figsetgrptitle{Subhalo 167392.}
\figsetplot{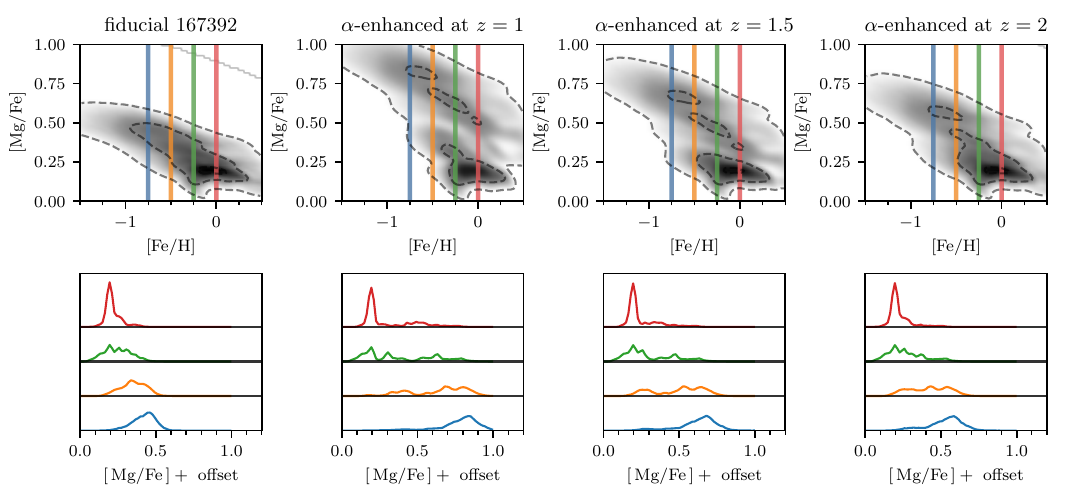}
\figsetgrpnote{Abundance plane for subhalo 167392.}
\figsetgrpend
\figsetgrpstart
\figsetgrpnum{8.6}
\figsetgrptitle{Subhalo 289388.}
\figsetplot{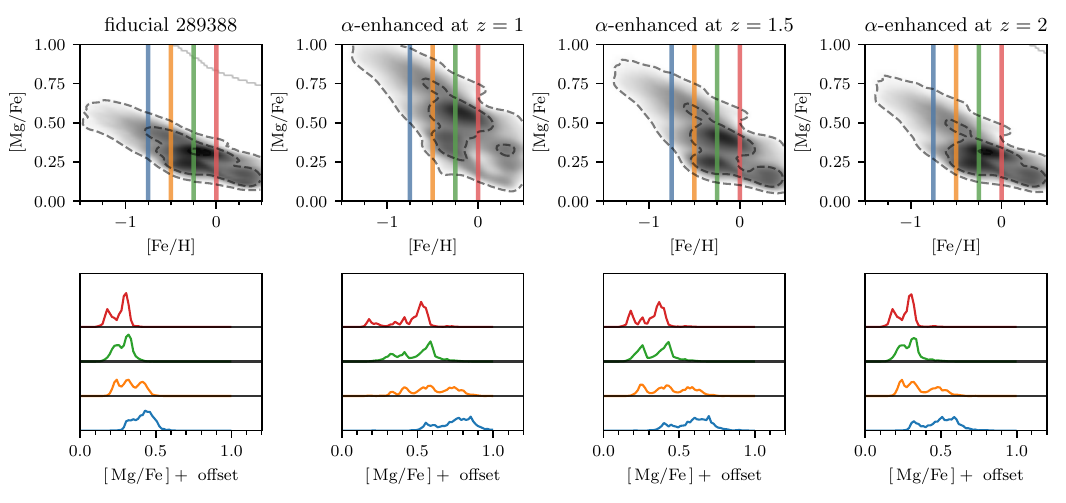}
\figsetgrpnote{Abundance plane for subhalo 289388.}
\figsetgrpend
\figsetgrpstart
\figsetgrpnum{8.7}
\figsetgrptitle{Subhalo 300903.}
\figsetplot{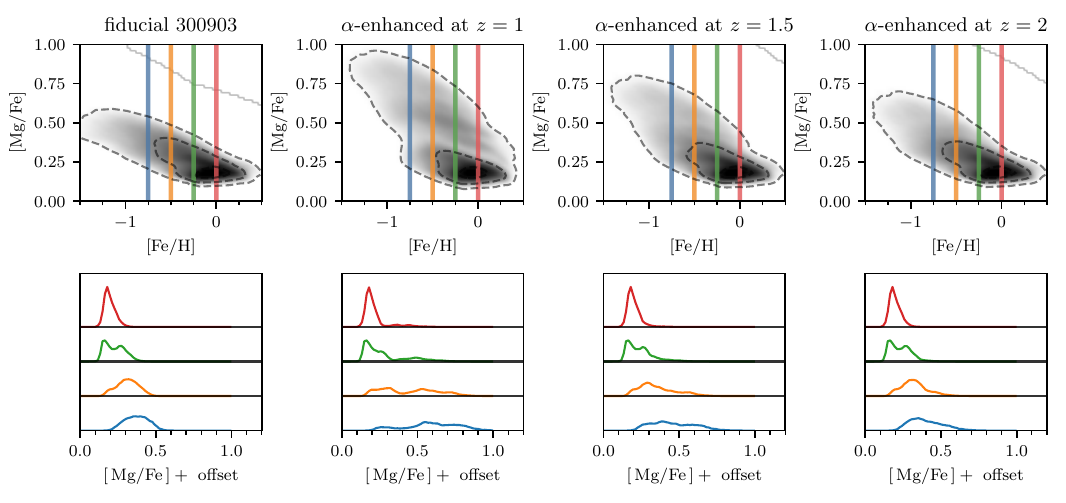}
\figsetgrpnote{Abundance plane for subhalo 300903.}
\figsetgrpend
\figsetgrpstart
\figsetgrpnum{8.8}
\figsetgrptitle{Subhalo 348901.}
\figsetplot{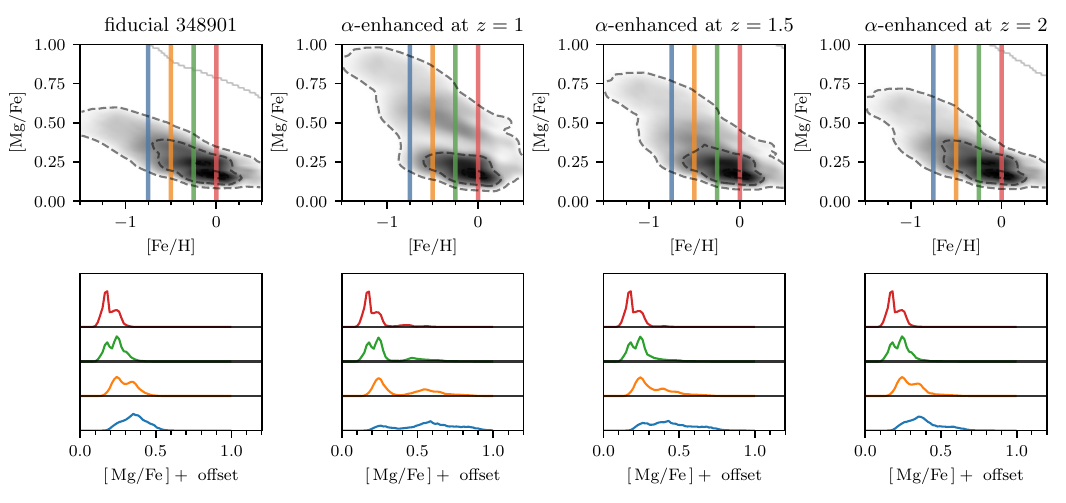}
\figsetgrpnote{Abundance plane for subhalo 348901.}
\figsetgrpend
\figsetgrpstart
\figsetgrpnum{8.9}
\figsetgrptitle{Subhalo 398784.}
\figsetplot{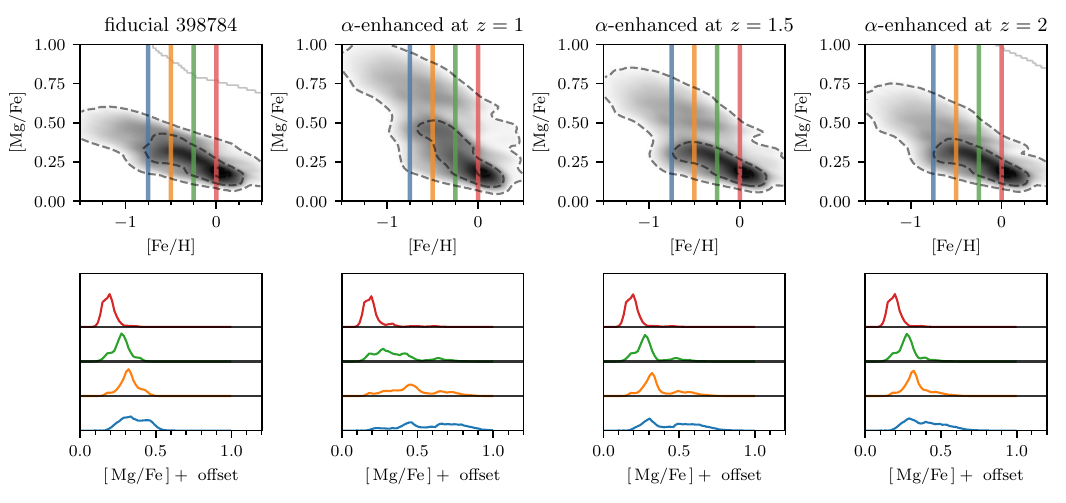}
\figsetgrpnote{Abundance plane for subhalo 398784.}
\figsetgrpend
\figsetgrpstart
\figsetgrpnum{8.10}
\figsetgrptitle{Subhalo 404818.}
\figsetplot{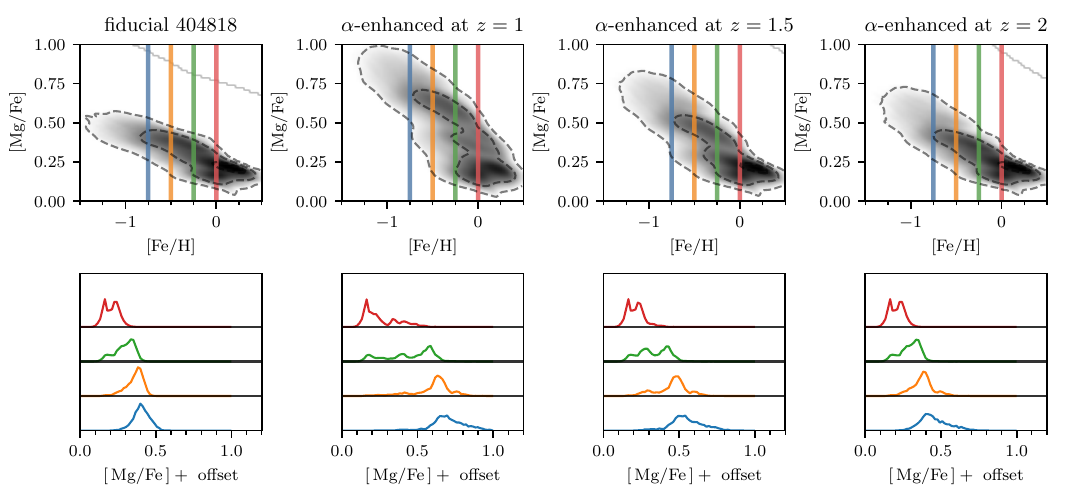}
\figsetgrpnote{Abundance plane for subhalo 404818.}
\figsetgrpend
\figsetgrpstart
\figsetgrpnum{8.11}
\figsetgrptitle{Subhalo 425719.}
\figsetplot{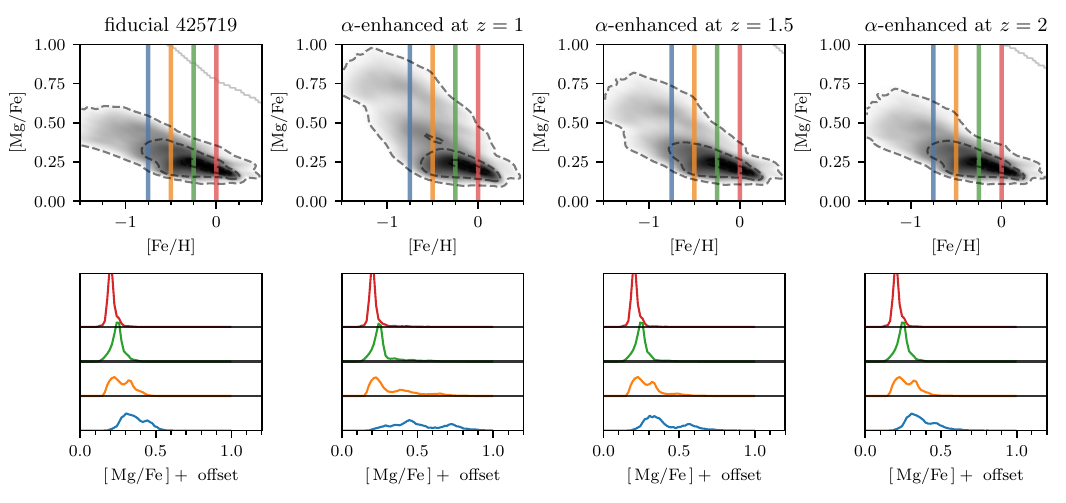}
\figsetgrpnote{Abundance plane for subhalo 425719.}
\figsetgrpend
\figsetgrpstart
\figsetgrpnum{8.12}
\figsetgrptitle{Subhalo 439099.}
\figsetplot{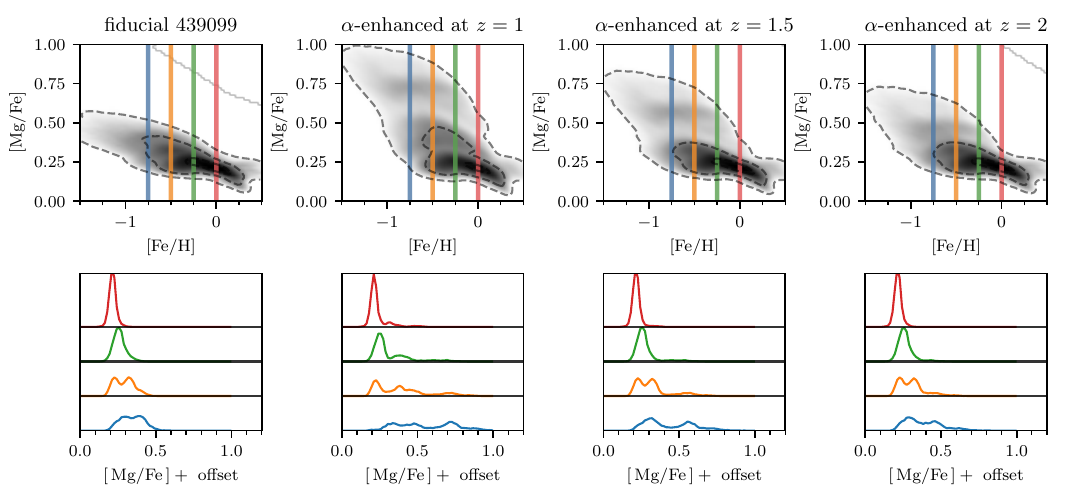}
\figsetgrpnote{Abundance plane for subhalo 439099.}
\figsetgrpend
\figsetgrpstart
\figsetgrpnum{8.13}
\figsetgrptitle{Subhalo 465255.}
\figsetplot{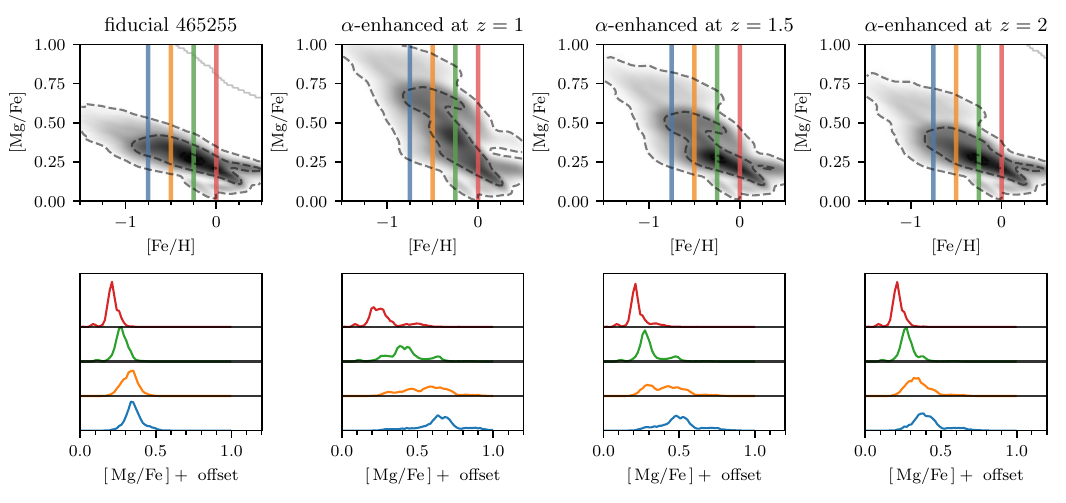}
\figsetgrpnote{Abundance plane for subhalo 465255.}
\figsetgrpend
\figsetgrpstart
\figsetgrpnum{8.14}
\figsetgrptitle{Subhalo 494709.}
\figsetplot{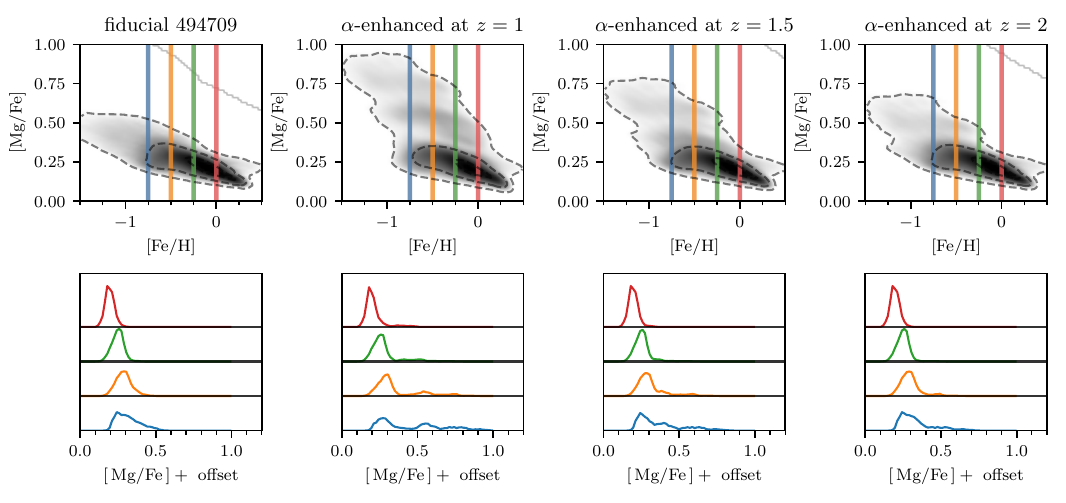}
\figsetgrpnote{Abundance plane for subhalo 494709.}
\figsetgrpend
\figsetgrpstart
\figsetgrpnum{8.15}
\figsetgrptitle{Subhalo 510273.}
\figsetplot{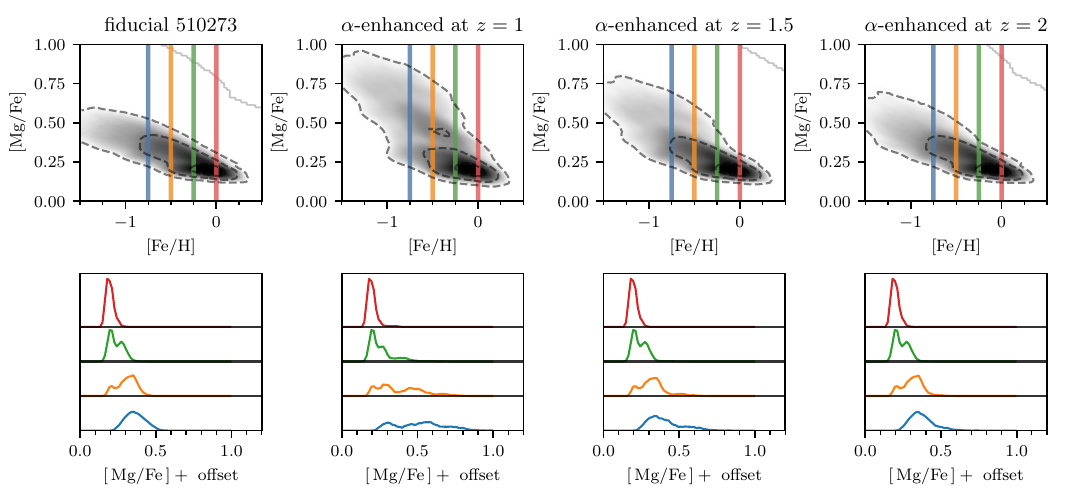}
\figsetgrpnote{Abundance plane for subhalo 510273.}
\figsetgrpend
\figsetgrpstart
\figsetgrpnum{8.16}
\figsetgrptitle{Subhalo 547293.}
\figsetplot{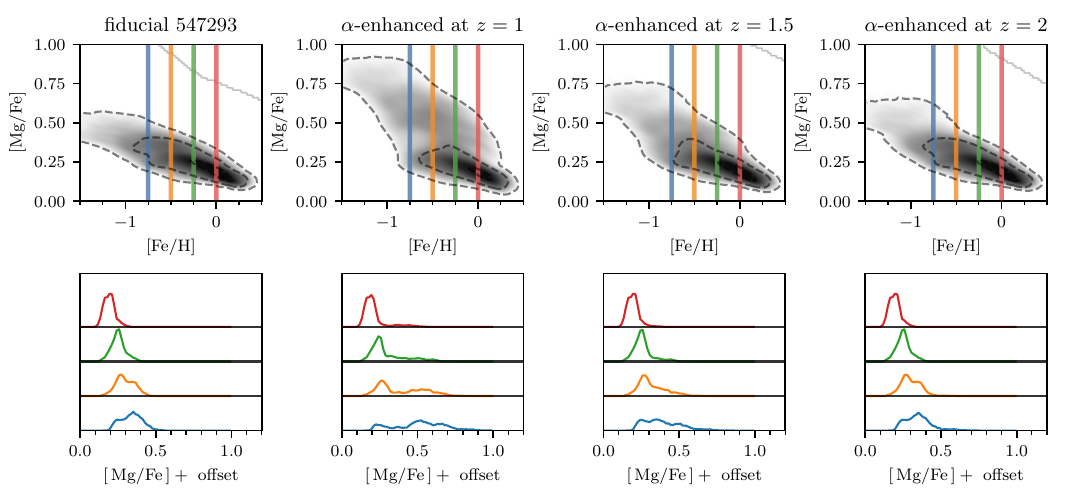}
\figsetgrpnote{Abundance plane for subhalo 547293.}
\figsetgrpend
\figsetgrpstart
\figsetgrpnum{8.17}
\figsetgrptitle{Subhalo 576516.}
\figsetplot{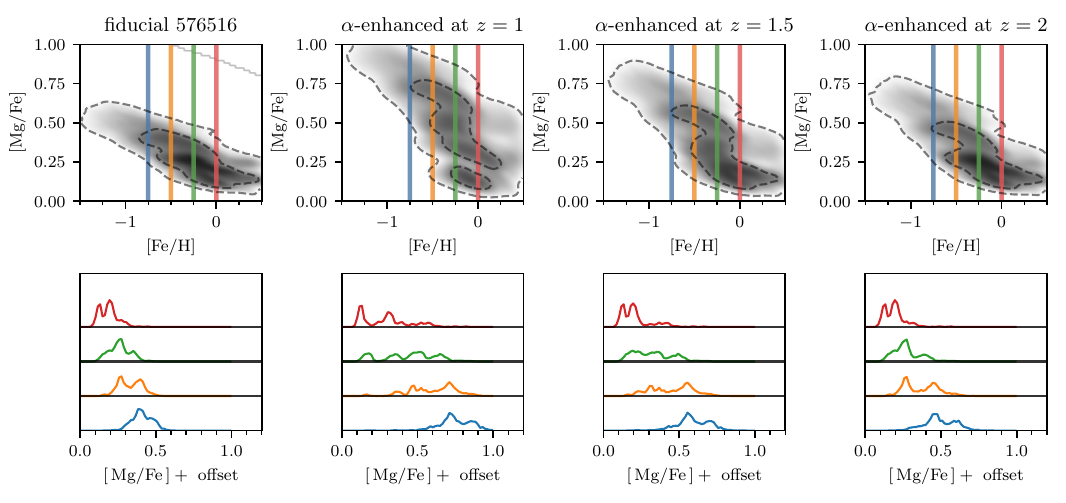}
\figsetgrpnote{Abundance plane for subhalo 576516.}
\figsetgrpend
\figsetend

\begin{figure*}
  \centering
  \includegraphics[width=\textwidth]{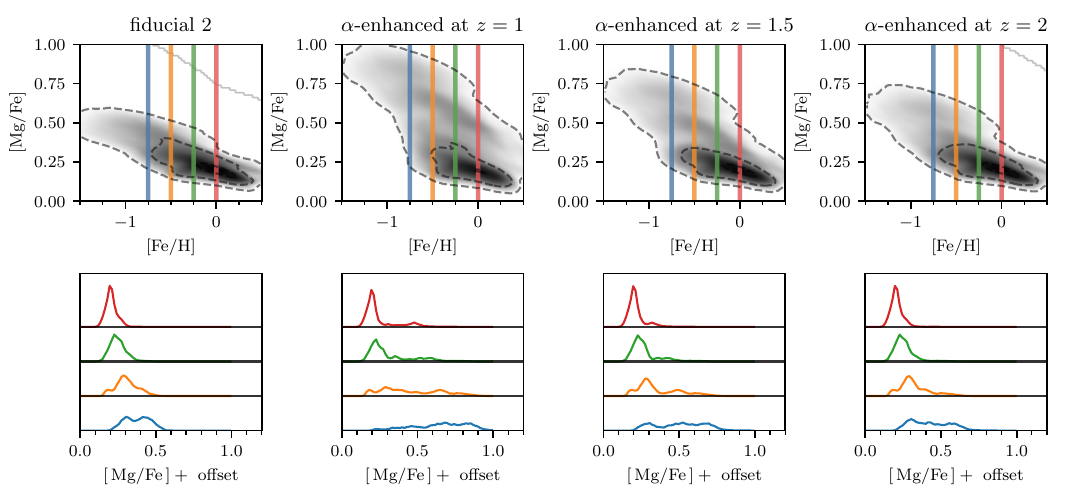}
  \caption{The same as Figure~\ref{fig:app0}, but for a random galaxy from our initial catalog.}
  \label{fig:app1}
\end{figure*}

\begin{figure*}
  \centering
  \includegraphics[width=\textwidth]{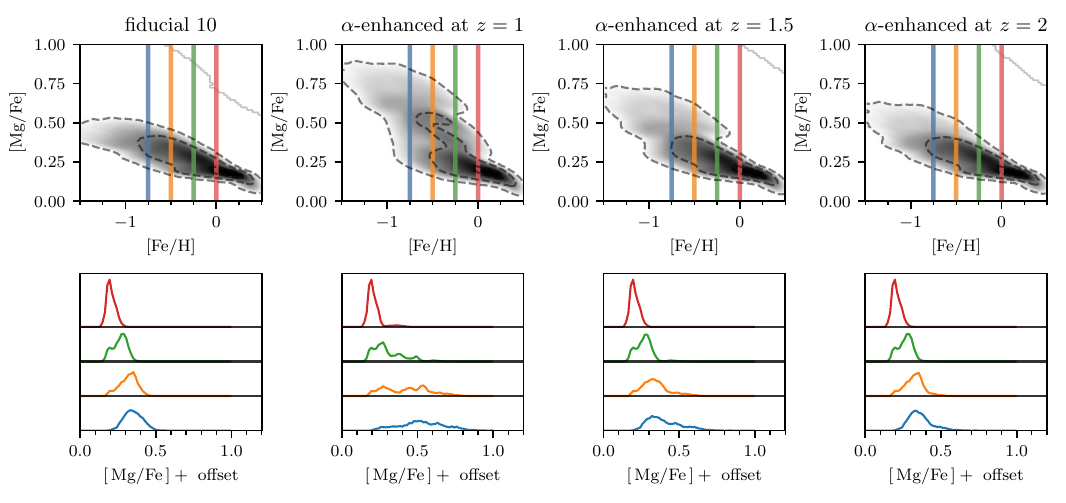}
  \caption{The same as Figure~\ref{fig:app0}, but for a random galaxy from our initial catalog.}
  \label{fig:app2}
\end{figure*}

\begin{figure*}
  \centering
  \includegraphics[width=\textwidth]{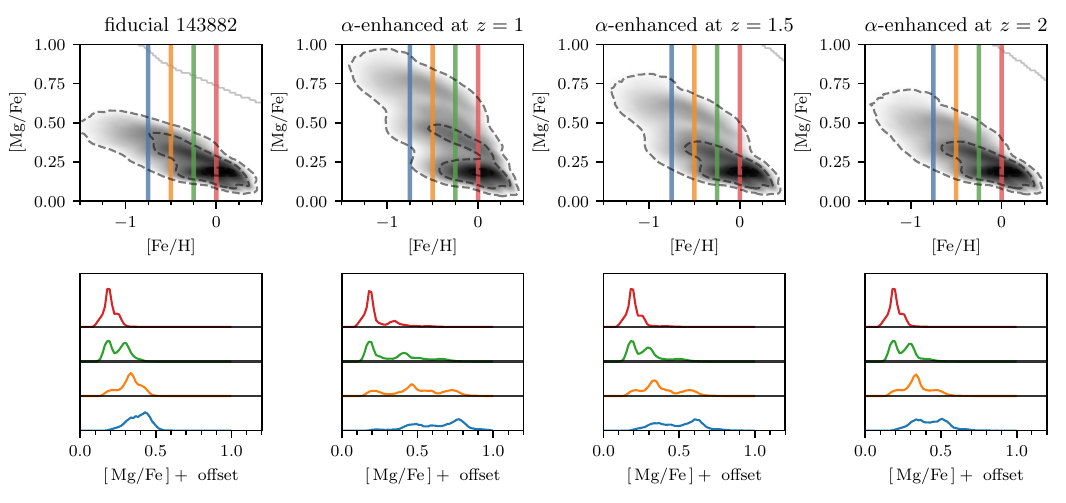}
  \caption{The same as Figure~\ref{fig:app0}, but for a random galaxy from our initial catalog.}
  \label{fig:app3}
\end{figure*}

\begin{figure*}
  \centering
  \includegraphics[width=\textwidth]{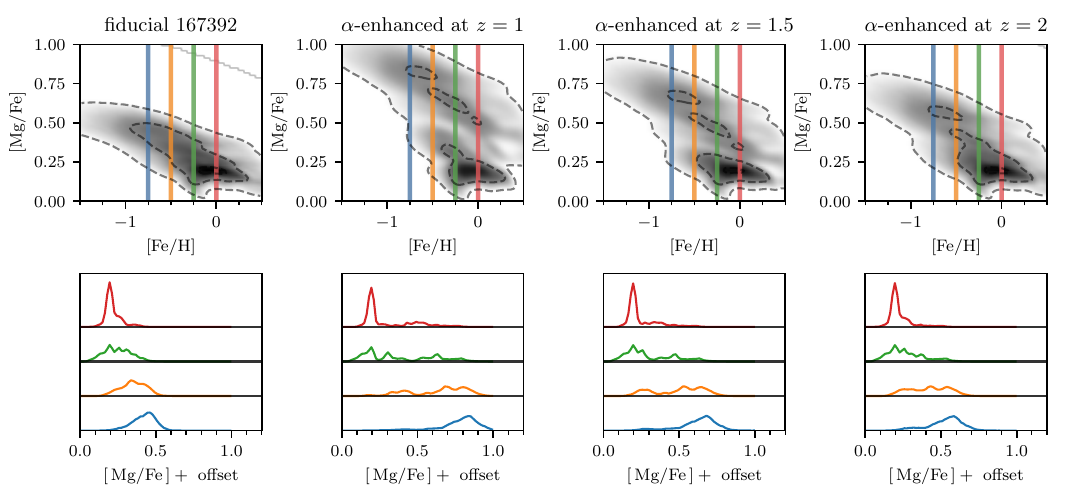}
  \caption{The same as Figure~\ref{fig:app0}, but for a random galaxy from our initial catalog.}
  \label{fig:app4}
\end{figure*}

\begin{figure*}
  \centering
  \includegraphics[width=\textwidth]{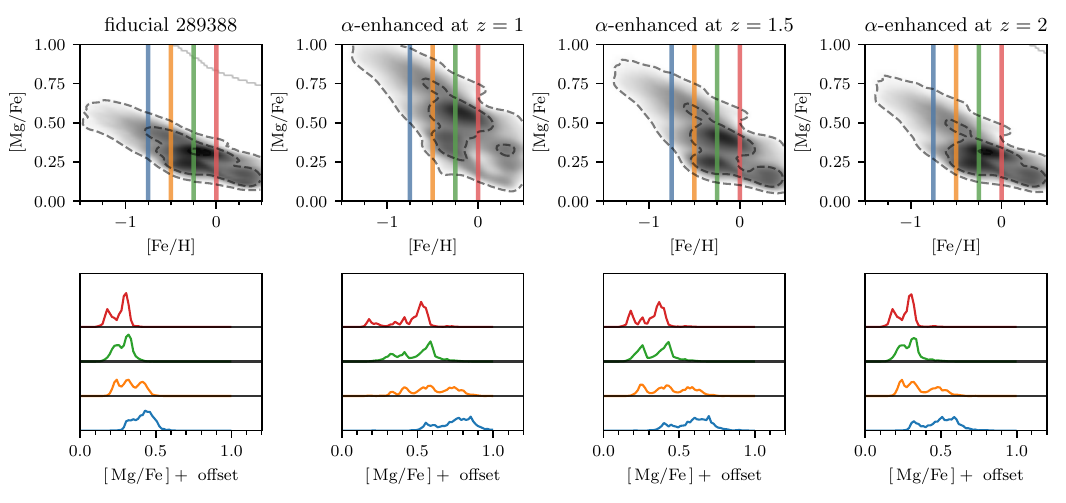}
  \caption{The same as Figure~\ref{fig:app0}, but for a random galaxy from our initial catalog.}
  \label{fig:app5}
\end{figure*}

\begin{figure*}
  \centering
  \includegraphics[width=\textwidth]{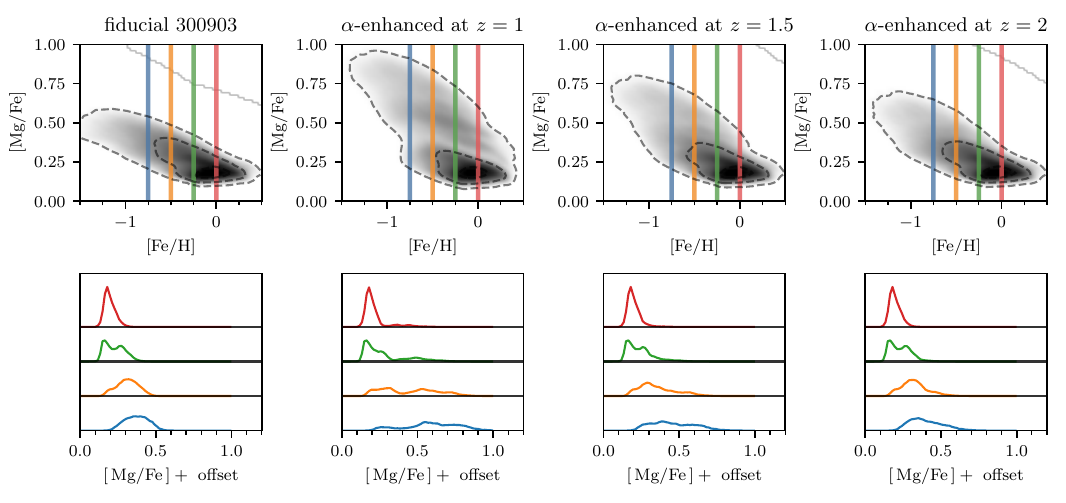}
  \caption{The same as Figure~\ref{fig:app0}, but for a random galaxy from our initial catalog.}
  \label{fig:app6}
\end{figure*}

\begin{figure*}
  \centering
  \includegraphics[width=\textwidth]{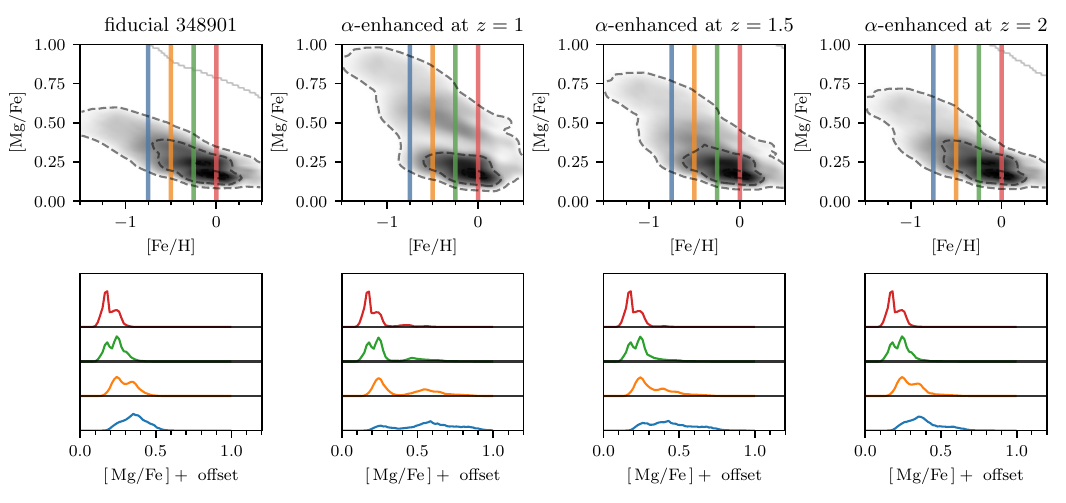}
  \caption{The same as Figure~\ref{fig:app0}, but for a random galaxy from our initial catalog.}
  \label{fig:app7}
\end{figure*}

\begin{figure*}
  \centering
  \includegraphics[width=\textwidth]{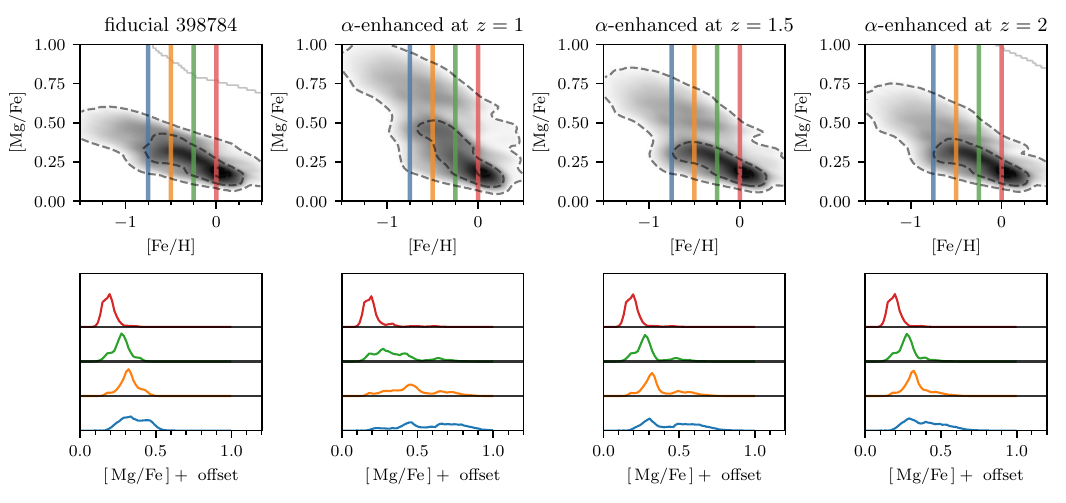}
  \caption{The same as Figure~\ref{fig:app0}, but for a random galaxy from our initial catalog.}
  \label{fig:app8}
\end{figure*}

\begin{figure*}
  \centering
  \includegraphics[width=\textwidth]{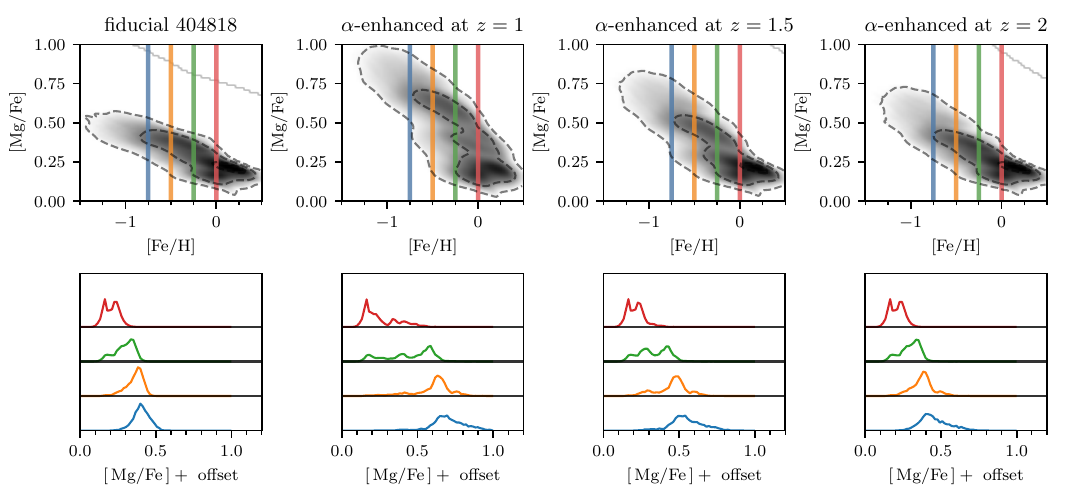}
  \caption{The same as Figure~\ref{fig:app0}, but for a random galaxy from our initial catalog.}
  \label{fig:app9}
\end{figure*}

\begin{figure*}
  \centering
  \includegraphics[width=\textwidth]{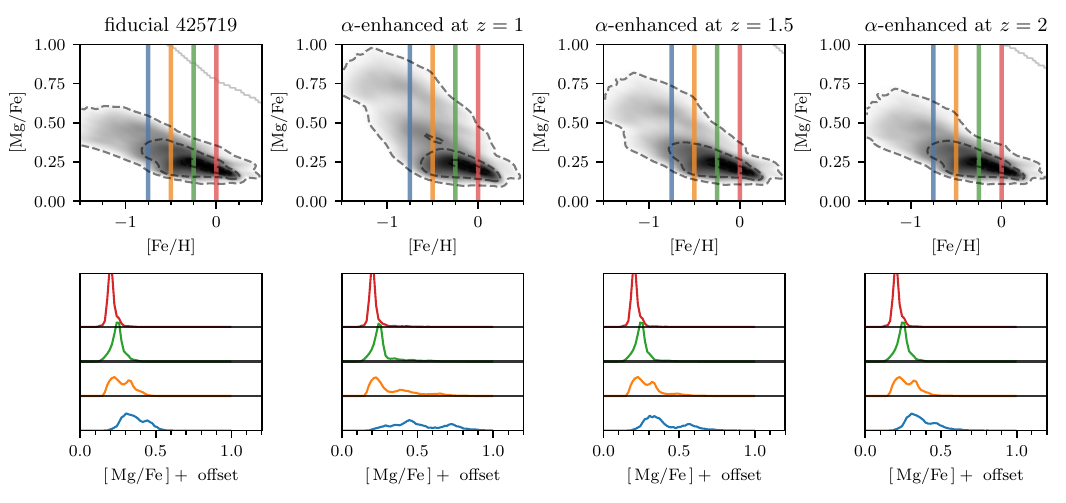}
  \caption{The same as Figure~\ref{fig:app0}, but for a random galaxy from our initial catalog.}
  \label{fig:app10}
\end{figure*}

\begin{figure*}
  \centering
  \includegraphics[width=\textwidth]{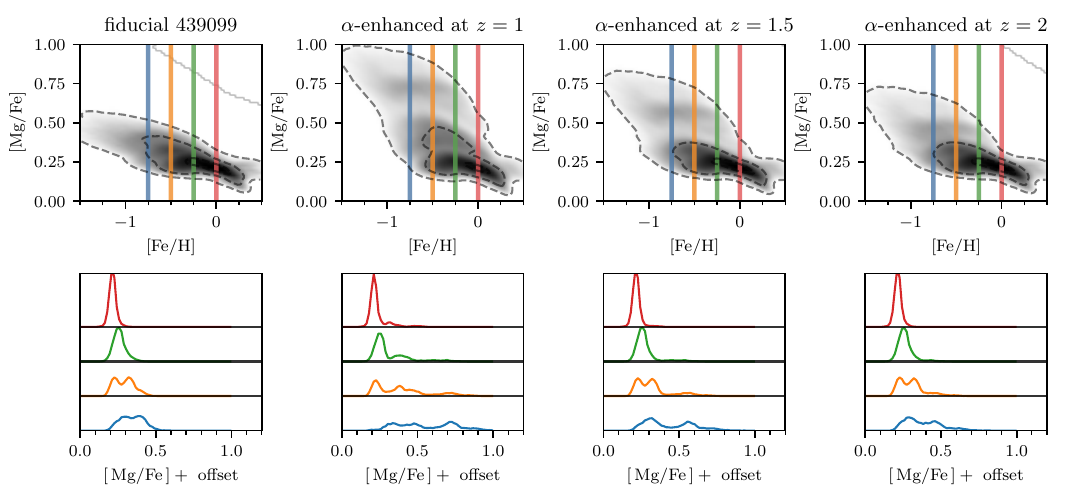}
  \caption{The same as Figure~\ref{fig:app0}, but for a random galaxy from our initial catalog.}
  \label{fig:app11}
\end{figure*}

\begin{figure*}
  \centering
  \includegraphics[width=\textwidth]{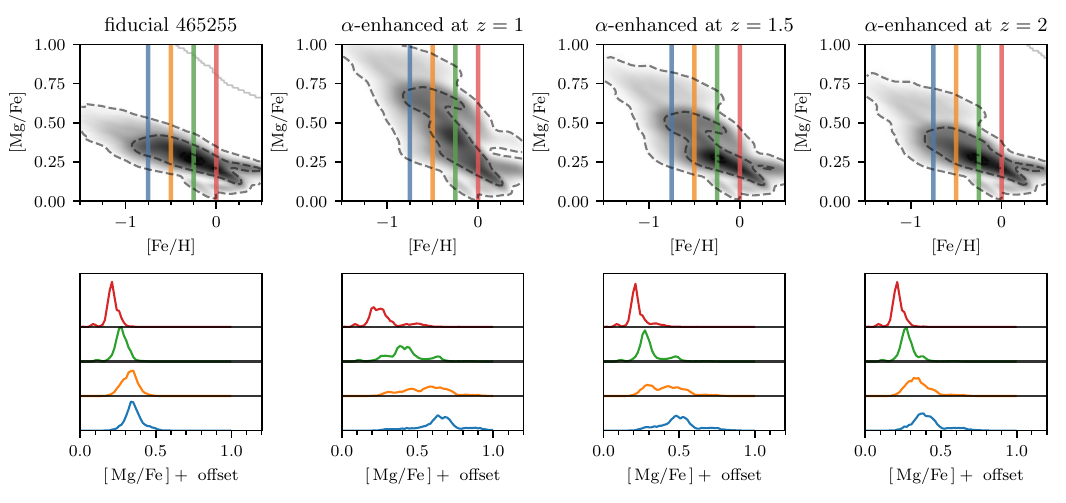}
  \caption{The same as Figure~\ref{fig:app0}, but for a random galaxy from our initial catalog.}
  \label{fig:app12}
\end{figure*}

\begin{figure*}
  \centering
  \includegraphics[width=\textwidth]{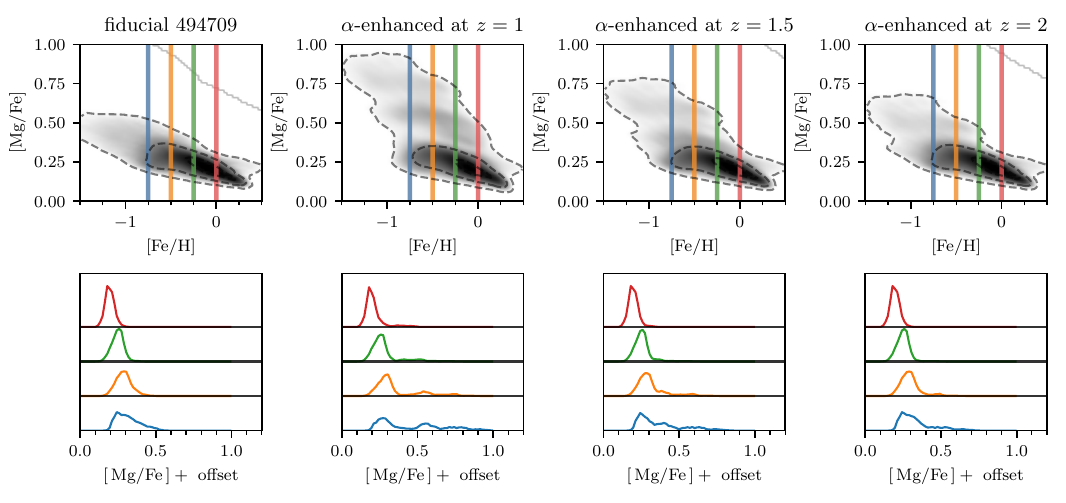}
  \caption{The same as Figure~\ref{fig:app0}, but for a random galaxy from our initial catalog.}
  \label{fig:app13}
\end{figure*}

\begin{figure*}
  \centering
  \includegraphics[width=\textwidth]{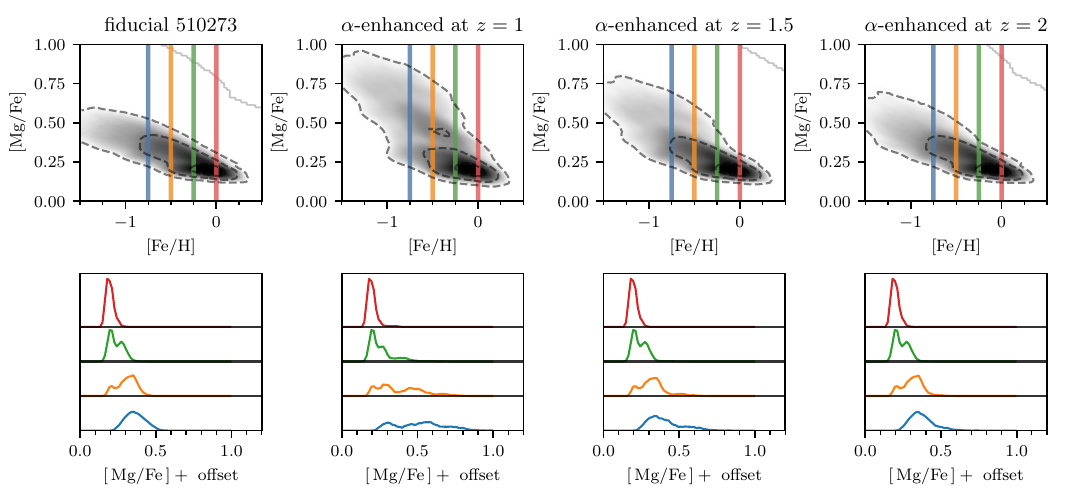}
  \caption{The same as Figure~\ref{fig:app0}, but for a random galaxy from our initial catalog.}
  \label{fig:app14}
\end{figure*}

\begin{figure*}
  \centering
  \includegraphics[width=\textwidth]{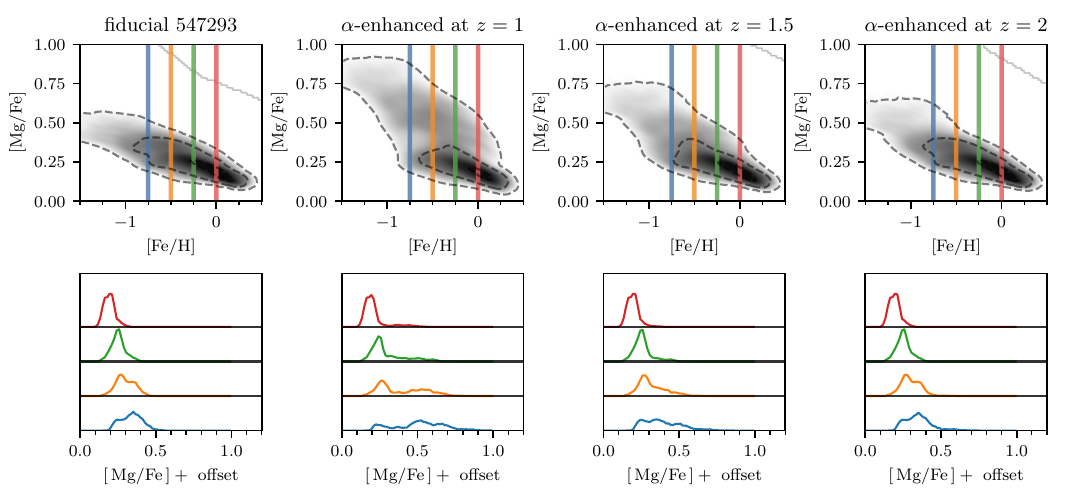}
  \caption{The same as Figure~\ref{fig:app0}, but for a random galaxy from our initial catalog.}
  \label{fig:app15}
\end{figure*}

\begin{figure*}
  \centering
  \includegraphics[width=\textwidth]{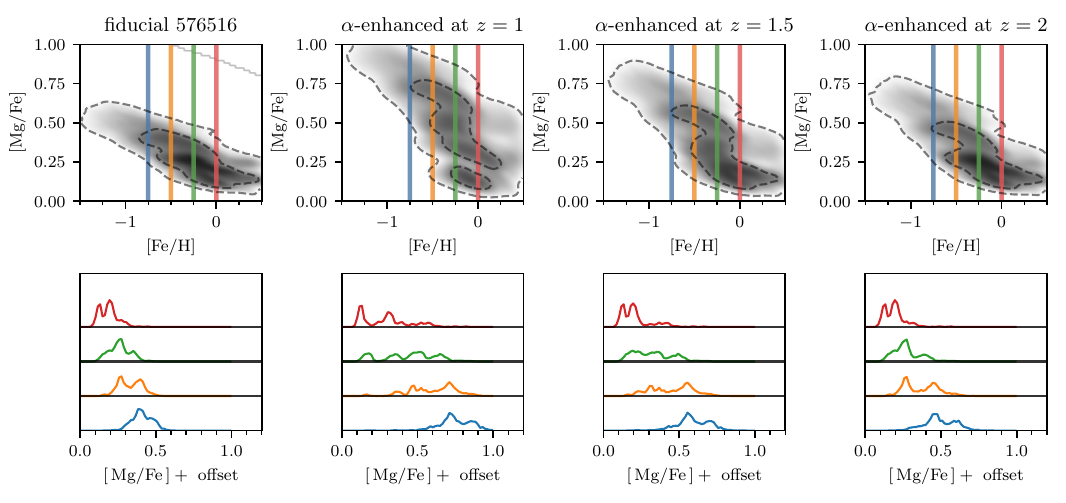}
  \caption{The same as Figure~\ref{fig:app0}, but for a random galaxy from our initial catalog.}
  \label{fig:app16}
\end{figure*}

\end{document}